\documentstyle[12pt]{article}
\input{epsf}
\renewcommand{\theequation}{\arabic{section}.\arabic{equation}}
\newcommand{\beq}{\begin{equation}}
\newcommand{\eeq}{\end{equation}}
\newcommand{\be}{\begin{eqnarray}}
\newcommand{\ee}{\end{eqnarray}}
\newcommand{\ba}{\begin{array}}
\newcommand{\ea}{\end{array}}
\newcommand{\bfv}{{\bf v}}
\newcommand{\bfp}{{\bf p}}
\newcommand{\bfgamma}{\mbox{\boldmath{$\gamma$}}}
\newcommand{\bftau}{\mbox{\boldmath{$\tau$}}}

\newcommand{\Tr}{{\rm Tr}\,}
\newcommand{\Sp}{{\rm Sp}\,}
\newcommand{\partialslash}{\partial\hspace{-.5em}/\hspace{.15em}}
\begin{document}
\rightline{NORDITA-97/24 P}
\rightline{RUB-TPII-3/97}
%
\vspace{1cm}
\begin{center}
{\bf\Large Unpolarized and polarized quark distributions in the
large--$N_c$ limit} \\[1cm]
{\bf\large D.I.~Diakonov$^{\rm a}$, V.Yu.~Petrov$^{\rm b}$,
P.V.~Pobylitsa$^{\rm b, c}$, M.V.~Polyakov$^{\rm b, c}$, and \\[.1cm]
C.~Weiss$^{\rm c}$} \\[.5cm]
{\it $^{a}$ NORDITA, Blegdamsvej 17, 2100 Copenhagen \O , Denmark \\
$^{b}$ Petersburg Nuclear Physics Institute, Gatchina,
St.Petersburg 188350, Russia \\
$^{c}$ Institut f\"ur Theoretische Physik II,
Ruhr--Universit\"at Bochum, D--44780 Bochum, Germany}
\end{center}
\vspace{1.5cm}
\begin{abstract}
\noindent
The isosinglet unpolarized and isovector polarized twist--2 quark
distributions of the nucleon at low normalization point are calculated
in the large--$N_c$ limit. The nucleon is described as a soliton of
the effective chiral theory. We derive the expressions for the
distribution functions in the large--$N_c$ limit starting from their
definition as numbers of partons carrying momentum fraction $x$
in the infinite momentum frame.  We develop a numerical
method for computation of the quark and antiquark distributions as
sums over the quark single--particle levels in the pion
field of the soliton.  The contributions of the discrete bound--state
level as well as the Dirac continuum are taken into account.  The
quark-- and antiquark distributions obtained explicitly satisfy all 
general requirements. Results are in reasonable agreement with
parametrizations of the data at low normalization point.
\end{abstract} 
\vspace{1.5cm} PACS: 13.60.Hb, 14.20.Dh, 12.38.Lg, 12.39.Ki, 11.15.Pg  
\newpage
\tableofcontents
\newpage
\section{Introduction}
\setcounter{equation}{0}
The evolution of parton distributions with $q^2$ in the asymptotic region
is well understood today, being governed by the renormalization group
equation of perturbative QCD. A complete description of experiments at
large $q^2$ requires, however, the knowledge of parton distributions in
the nucleon at some initial normalization point. Several sets of input 
distributions were determined by fits to the experimental data at large 
$q^2$ \cite{MRS95,CTEQ95, GRV95,GRSV96}. All these fits include antiquarks 
and gluons at a low normalization point.
\\
Recently, we have formulated an approach to calculate the twist--2 parton
distributions at low normalization point in the limit of a large number
of colors ($N_c$), where
the nucleon is described as a chiral soliton \cite{DPPPW96}.  At low
energies, QCD may be approximated by an effective theory whose degrees of
freedom are quarks with a dynamically generated mass, interacting with
pions, which appear as Goldstone bosons of the dynamically broken chiral
symmetry. The nucleon emerges as a classical soliton of
the pion field \cite{DPP88}.  This picture is known to give a successful
description of hadronic observables such as the nucleon mass, magnetic
moments, form factors {\em etc.} \cite{Review}. In \cite{DPPPW96} we have
shown that this approach possesses all necessary requisites for a
successful description of the leading--twist parton distributions of the
nucleon.  The normalization point of the distribution functions obtained in
this way is of the order of the ultraviolet cutoff of the effective chiral
theory, typically $\sim 600\, {\rm MeV}$.  Let us briefly summarize the main
characteristics of this description \cite{DPPPW96}:
\par
{\em i) Classification of quark distributions in the large $N_c$--limit.}
In the large--$N_c$ limit the quark distributions are concentrated at
values of $x\sim 1/N_c$. Combining this fact with the known large
$N_c$--behavior of the integrals of the distributions over $x$, one
infers that the quark distributions in the large $N_c$-- limit can
be divided in ``large'' and ``small'' ones. The leading distributions
are the isosinglet unpolarized and isosinglet polarized distributions,
which are of the form
\be
D^{\rm large}(x) &\sim& N_c^2 \rho (N_c x) ,
\ee
where $\rho (y)$ is a stable function in the large $N_c$--limit, which
depends on the particular distribution considered. The isovector unpolarized
and isosinglet polarized distributions appear only in the
next--to--leading order of the $1/N_c$--expansion, and are of the form
\be
D^{\rm small}(x) &\sim& N_c \rho (N_c x) .
\ee
\par
{\em ii) Sum rules and antiquark distributions.}
The chiral soliton model is a field--theoretic description of
the nucleon, which preserves all general requirements on parton
distributions. In particular, the standard sum rules for parton
distributions and their positivity properties are satisfied automatically
within the model. Also, a consistent description of the antiquark
distributions can be achieved in this approach.
\par
{\em iii) Parametric smallness of the gluon distribution.}
When working with the effective chiral theory, it is implied that the
ratio of the dynamical quark mass, $M$, to the UV cutoff, $\Lambda$
(not to be confused with the QCD scale parameter, $\Lambda_{\rm
QCD}$), is parametrically small. For $M/\Lambda \ll 1$, the quark
distributions computed in the effective theory may be identified with
the ``current'' quark distributions of QCD. The gluon distribution is
zero at this level, more precisely, it is $O(M^2 /\Lambda^2 )$. For
finite $M/\Lambda$, the quark distributions computed in the effective
theory should be interpreted as distributions of ``constituent''
quarks --- objects which themselves have a substructure in terms of
QCD partons.  The gluon distribution inside these objects could in
principle be recovered from the effective theory if one knew the
precise way how the UV cutoff arises as a result of integration over
the original QCD degrees of freedom.  These statements can be made
more precise in the framework of the instanton vacuum, which on one
hand allows to derive the effective chiral theory, on the other hand
can be used to evaluate the gluon distribution directly, using the
method developed in \cite{DPW96}. One finds that the gluon
distribution is suppressed relative to the quark distributions by a
factor of the packing fraction of the instanton medium \cite{DPPPW96}.
\par
In this paper, we study the properties of the $N_c$--leading quark and
antiquark distributions, namely the isosinglet unpolarized and
isovector polarized distributions, in the approach formulated in
\cite{DPPPW96}.  First, we rederive the basic formulas for the parton
distributions in the effective chiral theory in a new way. In
\cite{DPPPW96} these formulae were obtained from the exact QCD
expressions for the parton distributions as matrix elements of quark
bilinears with a light--like separation \cite{Jaffe,AEL95}. In this
paper we take the original Feynman point of view \cite{Feynman} that
parton distributions are given by the number of partons carrying a
fraction $x$ of the nucleon momentum in the nucleon infinite--momentum
frame.  Despite the apparent difference in wording we show here that
the two definitions are, in fact, equivalent and lead to identical
working formulae for computing parton distributions. We think it is
remarkable that the actual equivalence of the two well-known
definitions can explicitly be demonstrated within this
field--theoretical model of the nucleon.  The deep reason for the
equivalence is that the main hypothesis of the Feynman parton model,
namely that partons transverse momenta do not grow with $q^2$
\cite{Feynman}, is satisfied in the model under consideration.
\par
Second, we investigate the influence of the ultraviolet cutoff of the
effective chiral theory on the distribution functions. This not only
includes the asymptotic dependence on the cutoff parameter ({\em
i.e.}, the UV divergences), but also, and more importantly, the
dependence on the regularization scheme adopted to make the
distributions finite.  It is crucial that the regularization method
not violate the completeness of quark single--particle states in the
background pion field, in order to preserve causality, that is, the
anticommutation relations of quark fields at space--like separations.
A regularization which meets this requirement is, for example, the
Pauli--Villars subtraction. We show explicitly that it leads to quark
and antiquark distribution functions satisfying all general
requirements, such as rapid decrease for large $x$, uniform
logarithmic dependence on the cutoff, and positivity. On the other
hand, regularization methods based on an energy cutoff, such as the
popular proper--time regularization of the determinant, violate
causality and lead to unacceptable results for the distribution
functions.
\par
Third, we develop a numerical method for exact computation of the
quark and antiquark distribution functions as sums over quark
single--particle levels in the background pion field. (In
\cite{DPPPW96} the polarized distributions were estimated using an
approximation, the so-called interpolation formula).  Since the quark
distribution functions are given by matrix elements of products of
quark fields at finite time-- and space separations, their computation
requires evaluation of functional traces of the single--particle
energy and momentum operator, which is in general a difficult
problem. Using the finite--basis method of \cite{KR84}, we formulate a
reliable and efficient numerical procedure to compute the
Pauli--Villars regularized quark and antiquark distributions. The
method takes into account the contributions of the discrete
bound--state level as well as the Dirac continuum of quarks.
\par
Finally, using the methods developed in this paper, we compute the
isosinglet unpolarized and isovector polarized distribution functions
and discuss the results. In principle, the resulting distributions
should be taken as the starting point for perturbative evolution and
be compared with structure function data at large $q^2$. The evolution
of the calculated distributions and the comparison with the data will
be the subject of a separate investigation. Here, we restrict
ourselves to a comparison of the calculated distributions with the
parametrizations of the data at a low normalization point by Gl\"uck,
Reya {\em et al.}  \cite{GRV95,GRSV96}.
\section{Quark distribution functions in the large--$N_c$ limit}
\setcounter{equation}{0}
\subsection{The nucleon in the effective chiral theory}
In the large--$N_c$ limit, QCD becomes equivalent to an
effective theory of mesons, with baryons emerging as solitonic excitations
\cite{Witten,ANW}.  At low energies, the main guiding principle for
formulating this effective theory is the dynamical breaking of chiral
symmetry, which, in particular, results in the appearance of pions as
Goldstone bosons.  In the long--wavelength limit, the effective theory can
be expressed in the form of the chiral lagrangian of the pion field, whose
structure is basically determined by chiral symmetry. The minimal chirally
invariant interaction of quarks with Goldstone bosons is described by the
functional integral \cite{DE,DSW,DP86}
\be
\exp\left( i S_{\rm eff}[\pi (x) ] \right) &=&
\int D\psi D\bar\psi \; \exp\left(i\int d^4 x\,
\bar\psi(i\partialslash - M U^{\gamma_5})\psi\right) .
\label{effective_action}
\ee
Here, $\pi (x)$ is the pion field,
\be
U(x) &=& \exp\left[ i\pi^a(x)\tau^a\right] , \\
U^{\gamma_5}(x) &=& \exp\left[ i\pi^a(x)\tau^a\gamma_5\right] \; =
\; \frac{1+\gamma_5}2 U(x)
+ \frac{1-\gamma_5}2 U^\dagger (x).
\label{FI}
\ee
The quark field possesses a dynamical mass, $M$, due to chiral symmetry
breaking. It is understood that, generally, this theory of massive quarks
is valid up to an UV cutoff, $\Lambda \gg M$. The effective action
eq.(\ref{FI}) can be derived from the instanton vacuum, where the cutoff is
determined by the inverse instanton size, and the dynamical quark mass is
momentum dependent. In practice, rather than working with an explicitly
momentum--dependent quark mass, one usually takes a constant quark mass and
applies an UV cutoff to divergent quantities derived from
eq.(\ref{effective_action}), using some regularization scheme.
\par
In the effective chiral theory defined by eq.(\ref{effective_action}), the
nucleon is in the large--$N_c$ limit described by a static classical pion
field. In the nucleon rest frame it is of ``hedgehog'' form \cite{DPP88},
\be
U_c ({\bf x}) &=& \exp\left[ i ({\bf n} \cdot \bftau ) P(r) \right],
\hspace{1cm} r \; = \; |{\bf x}|,
\hspace{1cm} {\bf n} \; = \; \frac{{\bf x}}{r} .
\label{hedge}
\ee
Here, $P(r)$ is the profile function, with
$P(0) = -\pi$ and $P(r) \rightarrow 0$ for $r\rightarrow\infty$, which is
determined by minimizing the static energy of the pion field. Quarks are
described by means of one-particle wave functions, to be found from the
Dirac equation in the external pion field,
\be
\left(i\gamma^\mu\partial_\mu-MU^{\gamma_5}\right)\Psi_n({\bf x},t)
&=& 0,
\hspace{1.5cm}
\Psi_n({\bf x},t) \;\; = \;\; \exp(-i E_n t) \Phi_n ({\bf x}) .
\label{Dirac}
\ee
which can be written in Hamiltonian form,
\be
H\Phi_n &=& E_n\Phi_n ,
\hspace{1.5cm}
H \;\;
= \;\; -i\gamma^0 \gamma^k \partial_k + M \gamma^0 U^{\gamma_5} \,.
\label{H}
\ee
The spectrum of the one-particle Hamiltonian, $H$, contains a discrete
bound--state level. This level must be occupied by $N_c$ quarks to
have a state of unit baryon number. The nucleon mass is given by the
minimum of the bound--state energy and the aggregate energy of the negative
Dirac continuum, the energy of the free Dirac continuum subtracted
\cite{DPP88},
\be
M_N &=&  N_c E_{\rm lev}
\; + \; N_c \sum_{E_n<0} ( E_n - E_n^{(0)} ) .
\label{M_N}
\ee
($E_n^{(0)}$ denotes the energy levels of the vacuum Hamiltonian given
by eq.(\ref{H}) with $U = 1$.) It is understood that eq.(\ref{M_N}) is made
finite by some regularization method, to be discussed below.
\par
Nucleon states of definite spin-isospin and 3-momentum are obtained by
quantizing the rotational and translational zero modes of the soliton
(their contributions to the energy are $O(1/N_c )$), by integrating over
the corresponding collective coordinates with appropriate wave 
functions \cite{DPP88,Review}.
\subsection{Quark distribution functions in the effective chiral theory}
The simplest way to determine the quark distributions inside the nucleon is
to use the infinite momentum frame and calculate the number of partons
there \cite{Feynman}. It is well known that the infinite momentum helps to
separate quarks belonging to the nucleon from the vacuum ones, provided
that the transverse momenta of the particles are not growing with the
nucleon momentum \cite{Feynman}.  In our chiral quark soliton model this
condition is, of course, satisfied.
\par
More precisely, the quark distributions as functions of the Bjorken
variable, $x$, are, by definition, the number of (anti--) quarks whose
momentum, say, in the $z$ direction, is a fraction $x$ of the nucleon
momentum, $P_N$, in the infinite momentum frame where
\be
P_N &=& \frac{M_Nv}{\sqrt{1-v^2}},
\hspace{1.5em} v \; \rightarrow \; 1 .
\ee
The number of (anti--) quarks can be expressed through the nucleon matrix
element of the creation and annihilation operators, $a^+, a$ (for quarks),
and $b^+, b$ (for antiquarks). We define the quark and antiquark
distribution functions as
\be
D_i(x) &=&
\int\frac{d^3k}{(2\pi)^3}2\pi \delta\left( x - \frac{k^3}{P_N}\right)
\langle N_{{\bf v}}|a_i^+
({\bf k})a_i({\bf k})|N_{{\bf v}}\rangle , \\
\bar{D}_i(x) &=& \int\frac{d^3k}{(2\pi)^3}2\pi
\delta\left( x - \frac{k^3}{P_N}\right)
\langle N_{{\bf v}}|b_i^+(
{\bf k})b_i({\bf k})|N_{{\bf v}}\rangle  .
\label{distributions}
\ee
Here $i$ denotes the set of quantum numbers characterizing the quark, such
as flavor and polarization.
\par
In the infinite momentum frame it is possible to express these matrix
elements in terms of the quark field operator,
\beq
\psi({\bf x},t) = \int\frac{d^3k}{(2\pi)^3\sqrt{2k_0}}\sum_c\left[
a_c({\bf k})\exp(-i k\cdot x)u_c({\bf k}) +
b^+_c({\bf k})\exp(ik\cdot x)v_c({\bf k})
\right] ,
\label{field}
\eeq
where $u({\bf p})$, $v({\bf p})$ are the wave function of the free quarks
and antiquarks, normalized to $\bar{u}u = -\bar{v}v = 2M$, and $c = 1, 2$
denotes the polarization. Let us consider the Fourier transform of the
equal--time product of $\psi$ and $\psi^+$,
\be
\lefteqn{ \int\!
d^3x_1d^3x_2\exp[-i{\bf k}\cdot ({\bf x}_1-{\bf x}_2)]\psi^+({\bf x}_2,t)
\psi({\bf x}_1,t) } && \nonumber \\
&=& \sum_{c,c^\prime}\frac{1}{2k_0}\left[
a^+_c({\bf k})a_{c^\prime}({\bf k})u_c^*({\bf k})u_{c^\prime}({\bf k})
\right. \nonumber \\
&& +a^+_c({\bf k})b^+_{c^\prime}(-{\bf k})u_c^*({\bf k})v_{c^\prime}(-{\bf k})
\exp ( 2ik^0 t)
+
b_c(-{\bf k})a_{c^\prime}({\bf k})v^*_{c^\prime}(-{\bf k})u_c({\bf k})
\exp ( -2ik^0 t )
\nonumber \\[.5ex]
&& + \left.
b_c(-{\bf k})b^+_{c^\prime}(-{\bf k})v_c^*(-{\bf k})v_{c^\prime}(-{\bf k})
\right] .
\label{bilin}
\ee
Averaging this operator over the nucleon state in the infinite
momentum frame, with
$k^3 = xM_Nv/\sqrt{1-v^2}$, $v\rightarrow 1$, we get zero for all
terms on the r.h.s.\ but the first one.  Indeed, the probability to
find a correlated quark-antiquark pair with very large opposite
momenta (the second and third terms) in a fast moving nucleon goes to
zero as $v\rightarrow 1$.  Similarly, the probability to find
antiquarks moving in the opposite direction to the nucleon with large
longitudinal momenta goes to zero as $v\rightarrow 1$ (the fourth
term). To be more precise, these matrix elements decrease with the
nucleon momentum and can, in principle, contribute to the structure
functions of non-leading twists. However, we are interested in the
leading--twist distribution functions, and can therefore neglect all
terms in eq.(\ref{bilin}) except the first one.  The distribution functions
can thus be expressed in terms of equal-time products of the field
operators as
\be
D_i(x) &=& \int\!\frac{d^3k}{(2\pi)^3}
\delta\left( x - \frac{k^3}{P_N}\right)
\int\! d^3\!x_1 d^3\!x_2
\exp \left[ -i{\bf k}\cdot ({\bf x}_1-{\bf x}_2) \right] \nonumber \\
&& \times
\langle N_{{\bf v}}| \bar\psi({\bf x}_2, t)
\Gamma_i\psi({\bf x}_1, t)|N_{{\bf v}}\rangle ,
\label{quark} \\
\bar{D}_i(x) &=&
\int\!\frac{d^3k}{(2\pi)^3}
\delta\left( x - \frac{k^3}{P_N} \right)
\int\! d^3\!x_1 d^3\!x_2
\exp \left[ -i{\bf k}\cdot ({\bf x}_1-{\bf x}_2) \right]
\nonumber \\
&& \times
\langle N_{{\bf v}}|{\rm Tr}\left[ \Gamma_i
\psi({\bf x}_2,t)\bar\psi({\bf x}_1,t)\right] |N_{{\bf v}}\rangle .
\label{antiquark}
\ee
The flavor and spin matrices, $\Gamma_i$, depend on the particular
distribution one is interested in. For example, for the number of partons
polarized along or against the direction of the nucleon velocity one should
use
\beq
\Gamma_i = \gamma^0 \frac{1 \pm \gamma_5}{2} ,
\eeq
respectively.
\par
In order to evaluate the equal--time matrix elements in the nucleon state
in eqs.(\ref{quark}, \ref{antiquark}) we consider the more general matrix
element of the {\it time--ordered} product of quark fields.  In the
effective chiral theory, this nucleon matrix element can directly be
computed as a functional integral over the quark and pion field with the
effective action, eq.(\ref{effective_action}). At $N_c\rightarrow\infty$,
the integral can be performed using the saddle point method
\cite{DPP88}. One finds that
\beq
-i\langle N_{{\bf v}}| {\rm T}\, \left\{\psi({\bf x}_1,t_1)
\bar{\psi}({\bf x}_2,t_2)
\right\}|N_{{\bf v}}\rangle
= G_F(x_1, x_2) ,
\label{DefinedFeynman}
\eeq
where $G_F (x_1, x_2)$ is the Feynman Green function in the background
pion field corresponding to the moving nucleon.  This saddle point
solution of the classical equations of motion can easily be
constructed. Indeed, since the saddle point equations are relativistically
invariant, it is evident that the pion field is of the form
\beq
\hat{U}_c ({\bf x}, t)
= U_c \left(\frac{{\bf x}-{\bf v}t}{\sqrt{1-v^2}} \right) ,
\label{meanfield}
\eeq
where $U_c ({\bf x})$ is the stationary hedgehog pion mean field in the
nucleon rest frame. The Feynman Green function, eq.(\ref{DefinedFeynman}), 
is thus determined as the solution of the inhomogeneous Dirac equation
\beq
\left[
i\gamma^\mu\frac{\partial}{\partial x_1^\mu}
- M \hat{U}^{\gamma_5}({\bf x}_1 , t_1 )
\right] G_F({\bf x}_1 , t_1, {\bf x}_2 , t_2 ) = \delta^{(4)}(x_1 - x_2),
\label{DiracMoving}
\eeq
\beq
\hat{U}^{\gamma_5}({\bf x},t) = \frac{1+\gamma_5}2 \hat{U}_c ({\bf x}, t)
+ \frac{1-\gamma_5}2 \hat{U}_c^\dagger ({\bf x}, t) .
\eeq
With eq.(\ref{DefinedFeynman}), the distribution functions, 
eqs.(\ref{quark}, \ref{antiquark}), can now be written in terms of the
Feynman Green function as limits at $t_2\rightarrow t_1\pm 0$,
\be
D_i (x) &=& -i N_c\int\!\frac{d^3k}{(2\pi)^3}
\delta\left( x - \frac{k^3}{P_N} \right)
\nonumber \\
&& \times \int\! d^3\!x_1 d^3\!x_2
\exp\left[ -i{\bf k}\cdot ({\bf x}_1-{\bf x}_2) \right] {\rm Tr} \left[
\Gamma_i G_F({\bf x}_1,t_1,{\bf x}_2,t_2) \right]_{t_2=t_1+0} ,
\label{quarkD} \\
\bar{D}_i (x)&=& i N_c\int\!\frac{d^3k}{(2\pi)^3}
\delta\left( x - \frac{k^3}{P_N} \right)
\nonumber \\
&& \times \int\! d^3\!x_1 d^3\!x_2
\exp\left[ -i{\bf k} \cdot 
({\bf x}_1-{\bf x}_2) \right] {\rm Tr}\left[ \Gamma_i
G_F({\bf x}_1,t_1,{\bf x}_2,t_2) \right]_{t_1=t_2+0} .
\label{antiquarkD}
\ee
\par
To compute the quark and antiquark distribution functions one needs an
explicit representation of the Feynman Green function in the
background pion field corresponding to the fast--moving nucleon. We
now want to demonstrate that this Green function can be expressed in
terms of the single--particle quark wave functions, $\Phi_n
({\bf x})$, and energy eigenvalues, $E_n$, in the nucleon rest frame,
eq.(\ref{Dirac}).  The quark eigenfunctions in the time--dependent pion
field, eq.(\ref{meanfield}), can be obtained from the ones in the
static pion field in the rest frame by a Lorentz transformation. We
can thus write a representation of the Feynman Green function as
\be
\lefteqn{G_F({\bf x}_1, t_1, {\bf x}_2, t_2)} && \nonumber \\
&=& -iS[{\bf v}] \left\{ \theta(t_1-t_2)\!\!
\sum_{\rm non-occup.}
\Phi_n({\bf x}^\prime_1)\bar{\Phi}_n({\bf x}^\prime_2)
\exp\left[ -iE_n(t^\prime_1-t^\prime_2) \right] -\right.
\nonumber \\
&& \left. -\theta(t_2-t_1)\!\!
\sum_{\rm occup.}
\Phi_n({\bf x}^\prime_1 )\bar{\Phi}_n({\bf x}^\prime_2 )
\exp\left[ -iE_n(t^\prime_1-t^\prime_2)\right]
\right\} S^{-1}[{\bf v}].
\label{GreenFeynman}
\ee
Here $t^\prime$ and ${\bf x}^\prime$ are the Lorentz transforms of the
coordinates,
\beq
{\bf x}_{1,2}^\prime=
\frac{{\bf x}_{1,2}-{\bf v}t_{1,2}}{\sqrt{1-{\bf v}^2}},\hspace{1.5cm}
t_{1,2}^\prime=
\frac{t_{1,2}-{\bf v}\cdot{\bf x}_{1,2}}{\sqrt{1-{\bf v}^2}},
\eeq
and $S[{\bf v}]$ is the Lorentz transformation matrix acting on the
quark spinor indices,
\beq
S[{\bf v}]=\exp\left(\frac i2\sigma_{03}\omega\right), \quad
\sigma_{\mu\nu}=\frac{i}{2} [\gamma_\mu,\gamma_\nu],\quad {\rm
tanh}(\omega)=v.
\label{sv}
\eeq
In the first term of eq.(\ref{GreenFeynman}) the summation goes over
non-occupied states, that is over the positive-energy Dirac continuum, in
the second term over occupied states, that is, over the negative continuum
and the discrete bound-state level.
\par
Let us prove that the Green function defined by eq.(\ref{GreenFeynman}) 
indeed satisfies the inhomogeneous Dirac equation, eq.(\ref{DiracMoving}).  
We first note that the Lorentz transformed single--particle wave functions,
$S[{\bf v}]\Phi_n ({\bf x}^\prime)\exp(-iE_n t^\prime)$, satisfy the
homogeneous Dirac equation, eq.(\ref{DiracMoving}) with r.h.s.\ equal to
zero, therefore so does the Green function, eq.(\ref{GreenFeynman}), at 
$t_1 \ne t_2$. At $t_1 = t_2$ the Green function has a discontinuity.
Taking the time derivative of eq.(\ref{GreenFeynman}) at $t_1=t_2$ we find
that the l.h.s.\ of eq.(\ref{DiracMoving}) is equal to 
$F({\bf x}_1, {\bf x}_2) \delta (t_1 - t_2)$, where
\beq
F({\bf x}_1, {\bf x}_2) = \sum_{\rm all} \exp\left[ -iE_n
{\bf v}({\bf x}^\prime_1-{\bf x}^\prime_2) \right]
\gamma^0S\Phi_n({\bf x}^\prime_1 )\bar{\Phi}_n({\bf x}^\prime_2 )
S^{-1} .
\label{Delta}
\eeq
This function should be equal to $\delta^{(3)}({\bf x}_1-{\bf x}_2)$
in order to satisfy the inhomogeneous Dirac equation, eq.(\ref{DiracMoving}),
for the Green function\footnote{In field theory, eq.(\ref{Delta})
represents the equal--time anticommutator,
$\left\{\psi({\bf x}_1,t),\;\psi^+({\bf x}_2,t)\right\}=\delta^{(3)}
({\bf x}_1-{\bf x}_2)$.}.
To prove this, we first convince ourselves that the sum, eq.(\ref{Delta}),
vanishes for ${\bf x}_1 \ne {\bf x}_2$, and then show that it is indeed a
delta function with coefficient unity.
\par
We introduce temporarily two moments of ``time'',
$\tilde{t}_1={\bf v}{\bf x}^\prime_1$ and
$\tilde{t}_2={\bf v}{\bf x}^\prime_2$, and note that at
$\tilde{t}_1>\tilde{t}_2$ eq.(\ref{Delta}) can be written via 
the ``retarded'' Green function,
\be
F({\bf x}_1,{\bf x}_2) &=& i\gamma^{0}S \; G_{\rm ret}(
{\bf x}^\prime_1,({\bf v}{\bf x}^\prime_1),
{\bf x}^\prime_2,({\bf v}{\bf x}^\prime_2)) \; S^{-1},
\nonumber \\
G_{\rm ret}({\bf x}^\prime_1,\tilde{t}_1,{\bf x}^\prime_2,\tilde{t}_2)
&=& -i\theta(\tilde{t}_1-\tilde{t}_2)
\sum_{\rm all} \bar{\Phi}_n({\bf x}^\prime_2)
\Phi_n({\bf x}^\prime_1)\exp\left[ -iE_n(\tilde{t}_1-\tilde{t}_2) \right] .
\label{G-ret-sum}
\ee
We need this function in the space--like region, since
$(\tilde{t}_1-\tilde{t}_2)^2-({\bf x}^\prime_1-{\bf x}^\prime_2)^2
=({\bf v}\cdot ({\bf x}^\prime_1-{\bf x}^\prime_2))^2-
({\bf x}^\prime_1-{\bf x}^\prime_2)^2<0$, at least when
${\bf x}_1\ne{\bf x}_2$. However, the retarded Green function is zero in
the space-like region. This is obvious from physical considerations: the
retarded Green function determines the evolution of a wave packet which at
$\tilde{t}_1=\tilde{t}_2$ is localized at
${\bf x}^\prime_1= {\bf x}^\prime_2$, which cannot reach the space-like
region. More formally, this can also be proved using the perturbation
expansion for the retarded Green function in the external pion field,
\be
\lefteqn{
G_{\rm ret}({\bf x}^\prime_1,\tilde{t}_1,{\bf x}^\prime_2,\tilde{t}_2)
\;\; = \;\;
G_{\rm
ret}^0({\bf x}^\prime_1,\tilde{t}_1,{\bf x}^\prime_2,\tilde{t}_2) }
\nonumber \\
&+& \int d^3x^\prime_3d\tilde{t}_3
G_{\rm ret}^0({\bf x}^\prime_1,\tilde{t}_1,{\bf x}^\prime_3,\tilde{t}_3)
MU^{\gamma_5}({\bf x}^\prime_3)
G_{\rm ret}^0({\bf x}^\prime_3,\tilde{t}_3,{\bf x}^\prime_2,\tilde{t}_2)
\;\; + \;\; \ldots
\ee
The free retarded Green function, $G^0_{\rm ret}$, is zero for space-like
separations, therefore the second term is non-zero only for
$\tilde{t}_1-\tilde{t}_3>|{\bf x}^\prime_1-{\bf x}^\prime_2|$ and for
$\tilde{t}_3-\tilde{t}_2>|{\bf x}^\prime_2-{\bf x}^\prime_3|$. This implies
that it is non-zero only for $\tilde{t}_1-\tilde{t}_2>
|{\bf x}^\prime_1-{\bf x}^\prime_3|+|{\bf x}^\prime_3-{\bf x}^\prime_2|>
|{\bf x}^\prime_1-{\bf x}^\prime_2|$.  This argument is easily generalized
to any term of the perturbation expansion.
\par
Thus, $F({\bf x}_1,{\bf x}_2)$ is zero for ${\bf x}_1\ne{\bf x}_2$. It
therefore must be proportional to a delta function in 
${\bf x}_1 - {\bf x}_2$ or its derivatives.  This ``point''
singularity can come only from states with large energies, $E_n$. For
such states, one can neglect the pion field and replace the wave
functions, $\Phi_n({\bf x})$, by the eigenfunction of the free
Hamiltonian. Saturating the sum in the r.h.s.\ of eq.(\ref{G-ret-sum})
by plane waves we obtain
\beq
F({\bf x}_1,{\bf x}_2)
=\delta^{(3)}({\bf x}_1-{\bf x}_2) \,.
\label{completeness}
\eeq
(In other words, the leading short--distance singularity of the Green
function in the background pion field is the same as that of the free Green
function.)  This completes the proof that eq.(\ref{GreenFeynman}) is a
representation of the Feynman Green function in the time--dependent 
background field, eq.(\ref{DiracMoving}).
\par
We now use the representation, eq.(\ref{GreenFeynman}), to express the
distribution functions directly through the quark wave functions in the
nucleon rest frame. Passing to the Fourier transforms of the quark wave
functions, integrating in eqs.(\ref{quarkD}, \ref{antiquarkD}) over 
${\bf x}_{1,2}$, and taking the limit $v\rightarrow 1$, we finally obtain
\be
D_i(x) &=& N_cM_N\!\!\sum_{\rm
occup.}\int\!\frac{d^3k}{(2\pi)^3}
\Phi_n^\dagger ({\bf k}) (1+\gamma^0\gamma^3) \gamma^0 \Gamma_i
\delta(k^3 + E_n - xM_N) \Phi_n ({\bf k}) ,
\label{FinalQuarks}
\\
\bar{D}_i(x) &=& N_cM_N\!\!\sum_{\rm
non-occup.}\int\!\frac{d^3k}{(2\pi)^3}
\Phi_n^\dagger ({\bf k}) (1+\gamma^0\gamma^3) \gamma^0 \Gamma_i
\delta(k^3 + E_n + xM_N) \Phi_n ({\bf k}) .
\nonumber \\
\label{FinalAnti}
\ee
These formulae represent the quark distributions as sums over occupied
states, the antiquark distributions as sums over non-occupied states.
We can also write an alternative representation for the distribution
functions, using the time--ordering opposite to the one in
eqs.(\ref{quarkD}, \ref{antiquarkD}), and the
fact that the discontinuity of the Feynman Green function is a space
$\delta$--function. In this case we get
\be
D_i(x) &=& -N_cM_N\!\!\sum_{\rm
non-occup.}\int\!\frac{d^3k}{(2\pi)^3}
\Phi_n^\dagger ({\bf k})(1+\gamma^0\gamma^3)\gamma^0 \Gamma_i 
\delta(k^3 + E_n - xM_N) \Phi_n ({\bf k}) ,
\nonumber \\
\label{EquivQuarks}
\\
\bar{D}_i(x) &=& -N_cM_N\!\!\sum_{\rm
occup.}\int\!\frac{d^3k}{(2\pi)^3}
\Phi_n^\dagger ({\bf k})(1+\gamma^0\gamma^3)\gamma^0 \Gamma_i
\delta(k^3 + E_n + xM_N) \Phi_n ({\bf k}) .
\label{EquivAnti}
\ee
In both eqs.(\ref{FinalQuarks}, \ref{FinalAnti}) and
eqs.(\ref{EquivQuarks}, \ref{EquivAnti}) it is understood that the
contribution of free quarks is to be subtracted. In the original
representation of the distribution functions through the
Feynman Green function, eqs.(\ref{quarkD}, \ref{antiquarkD}), this
means that the leading short--distance singularity of the Green
function in the background pion field is canceled by the one of
the free Green function, {\em cf.}\ eq.(\ref{completeness}). However,
eqs.(\ref{FinalQuarks}, \ref{FinalAnti}) and
eqs.(\ref{EquivQuarks}, \ref{EquivAnti}) still contain ultraviolet
divergences and have to be made finite by some regularization scheme,
to be discussed in Section 3.
\par
It is clear that the equivalence of the representations of the distribution
functions as sums over occupied and non-occupied states,
eqs.(\ref{FinalQuarks}, \ref{FinalAnti}) and 
eqs.(\ref{EquivQuarks}, \ref{EquivAnti}), 
is based on the completeness of quark states in the external pion field,
eq.(\ref{completeness}), which can in principle be violated by
ultraviolet regularization.  Fortunately,  it is possible to regularize
the theory in such a way that this important property is preserved, see
Section 3.
\par
We emphasize that formulae eqs.(\ref{FinalQuarks}, \ref{FinalAnti}) and
eqs.(\ref{EquivQuarks}, \ref{EquivAnti}) are identical to those which have
been derived in ref.\cite{DPPPW96} from a representation of the
distribution functions as nucleon matrix elements of quark bilinears
separated by a light--like distance. The above derivation is a
demonstration of the equivalence of the two definitions of parton
distributions.
\par
Finally, to get the unpolarized or polarized (anti--) quark distributions
corresponding to a nucleon state of definite spin and isospin, one has to
take the desired combinations of the basic expressions
eqs.(\ref{FinalQuarks}, \ref{FinalAnti}) and average them with the nucleon
spin--isospin wave function (the procedure is described in
\cite{DPPPW96}).  In this paper we consider the distributions which appear
in the leading order of the $1/N_c$--expansion, the isosinglet unpolarized
and the isovector polarized one. To obtain the isosinglet unpolarized
quark distribution we sum up the two polarizations in 
eq.(\ref{FinalQuarks}) and
average over flavor. One finds \cite{DPPPW96}
\be
[u(x) \, + \, d(x)]_{\rm occup.} &=&
N_c M_N
\!\!\sum\limits_{\scriptstyle n\atop \scriptstyle{\rm occup.}}\!
\int\!\frac{d^3k}{(2\pi)^3}
\Phi_n^\dagger ({\bf k})(1+\gamma^0\gamma^3) \delta(k^3 + E_n - xM_N)
\Phi_n ({\bf k}) . 
\nonumber \\
\label{singlet_occ} 
\ee
{From} eq.(\ref{EquivAnti}) one sees that the corresponding antiquark
distribution can also be written as a sum over occupied states; it is
given by the negative of the r.h.s.\ of eq.(\ref{singlet_occ}) at 
$x \rightarrow -x$.  Henceforth, we shall consider
eq.(\ref{singlet_occ}) as a function defined for both positive and
negative $x$, and understand that at negative $x$ it describes minus
the antiquark distribution.  Alternatively, we can use
eqs.(\ref{FinalAnti}, \ref{EquivQuarks}) to obtain a representation of
the isosinglet unpolarized distribution as
\be
\lefteqn{ [u(x) \, + \, d(x)]_{\rm non-occup.}} && \nonumber \\
&=& - N_c M_N
\!\!\sum\limits_{\scriptstyle n\atop \scriptstyle{\rm non-occup.}}\!
\int\!\frac{d^3k}{(2\pi)^3}
\Phi_n^\dagger ({\bf k})(1+\gamma^0\gamma^3) \delta(k^3 + E_n - xM_N)
\Phi_n ({\bf k}) ,
\label{singlet_nonocc} 
\ee
which, again, gives minus the antiquark distribution at negative $x$. 
In eq.(\ref{singlet_occ}) vacuum subtraction is understood 
for $x < 0$, in eq.(\ref{singlet_nonocc}) for $x > 0$.
\par
The analogous expression for the isovector polarized distribution is
\be
\lefteqn{[\Delta u(x) \, - \, \Delta d(x)]_{\rm occup.}} && \nonumber \\
&=& -\frac{1}{3} (2T_3)  N_c M_N
\!\!\sum\limits_{\scriptstyle n\atop \scriptstyle{\rm occup.}}\!
\int\!\frac{d^3k}{(2\pi)^3}
\Phi_n^\dagger ({\bf k}) \tau^3 (1+\gamma^0\gamma^3) \gamma_5
\delta(k^3 + E_n + xM_N) \Phi_n ({\bf k}) ,
\nonumber \\
\label{isovector_occ}
\ee
with the corresponding antiquark distribution given by the same
expression with $x \rightarrow -x$ (vacuum subtraction is again
understood).  Here, $2T_3 = \pm 1$ for proton and neutron,
respectively. The alternative representation as a sum over
non-occupied states analogous to eq.(\ref{singlet_nonocc}) can easily
be written down.
\par
The isovector unpolarized and isosinglet polarized quark distributions
vanish in the leading order of the $1/N_c$--expansion. They are non-zero
only after considering rotational corrections, {\em i.e.}, expanding to
first order in the soliton angular velocity, which is $O(1/N_c )$, and are
expressed as double sums over single--particle levels.  We shall not
consider them in this paper. We note, however, that the techniques
developed in Sections 3 and 4 can readily be generalized to analyze also
these ``small'' distributions.
\section{Ultraviolet divergences and regularization}
\setcounter{equation}{0}
The expressions for the quark distribution functions
derived in the previous section are ultraviolet
divergent and require regularization. To be able to compute
the distribution functions using the effective chiral theory
we must ensure that the
ultraviolet regularization does not lead to violation of any of their
fundamental properties. We want to show now that
regularization by a Pauli--Villars subtraction, which preserves the
completeness of the quark single--particle states in the chiral
soliton, leads to regularized quark and antiquark distributions
satisfying all general requirements.
It was noted in Section 2 that the equivalence of the representations
of the distribution functions as sums over occupied and non-occupied
states relies on the completeness of the single--particle states.  We
shall see below that, with Pauli--Villars regularization, this
equivalence is preserved for the regularized distributions.  On the
other hand, regularizations based on an energy cutoff not only destroy 
the equivalence of summation over occupied and non-occupied states, but
lead also to other, related, unphysical features.
\par
The divergent contribution in eq.(\ref{singlet_occ})
comes from the eigenstates of the Dirac Hamiltonian
with large energy, $|E_n|$. One may think therefore that a natural way to
regularize this divergence is simply to cut the contributions
of states with $|E_n|$ larger than some cutoff, $\omega_0$:
\be
[u(x) + d(x)]_{\rm occup.}^{\omega_0} &=&  N_c M_N
\int\limits_{-\omega_0}^{ E_{\rm lev}+0} d\omega \;
\Sp \left[ \delta(\omega-H) \delta(\omega -x M_N + p^3 )
(1 + \gamma^0 \gamma^3 ) \right]
\nonumber \\
&&
- (H\to H_0) .
\label{energy-cutoff-occupied}
\ee
(Here we have written the sum over states, eq.(\ref{singlet_occ}), as 
an integral over energy, the integrand being a functional
trace involving the Hamiltonian and single--particle momentum
operator, $p^3$ \cite{DPPPW96}. This form is useful for
investigating the ultraviolet asymptotics.)  In fact, we shall see
below that this regularization is unphysical, leading to a number of
problems which are easily cured by turning to the Pauli--Villars
regularization.  However, two reasons force us to devote some time to
the distribution function with the energy cutoff,
eq.(\ref{energy-cutoff-occupied}). First, in order to see explicitly
that the Pauli--Villars subtraction cancels all non-physical effects,
one has to understand precisely what is to be canceled. Second, our 
numerical method for computing the distribution function (see Section 4)
involves an energy cutoff in the intermediate stages of the
calculation (taken to infinity at the end), so it is essential to know
the asymptotic properties of the distributions with energy cutoff.
\par
The ultraviolet divergence of eq.(\ref{energy-cutoff-occupied}) can be
derived using the technique developed in \cite{DPPPW96}. One replaces
the $\delta$--function by the imaginary part (discontinuity) of the
quark propagator in the background pion field,
\be
\delta (\omega - H) &=& \frac{1}{2\pi i } \left(
\frac{1}{\omega - H - i0} - \frac{1}{\omega - H + i0} \right) .
\label{delta_to_propagator}
\ee
Writing the quark propagator in the form
\be
\frac{1}{\omega - H} &=& \frac{\omega + H}{\omega^2 - H^2}
\;\; = \;\; \frac{\omega + \gamma^0 (-i\gamma^k\partial_k + M U^{\gamma_5})}
{\omega^2 + \partial_k^2 - M^2 - iM \gamma^k (\partial_k U^{\gamma_5})} ,
\label{propagator_squared}
\ee
one can expand in derivatives of the soliton field. One finds that
only the first term in the derivative expansion is divergent.  In this
way one easily obtains the logarithmic divergence of the distribution
function, eq.(\ref{energy-cutoff-occupied}), in the limit of large
energy cutoff,
\be
\lefteqn{ [ u(x) + d(x)]_{\rm occup.}^{\omega_0}} && \nonumber \\
&\sim& N_c M_N M^2 \log \frac{\omega_0}{M}
\mbox{sign}(x) \frac{1}{4\pi^2} 
\int \frac{d^3 {\bf k}}{(2\pi)^3}
\theta( k^3 - M_N |x| ) \;
\Tr \left[ \tilde U ({\bf k}) [\tilde U ({\bf k})]^\dagger \right] ,
\nonumber \\
&& (\omega_0 \rightarrow \infty ) .
\label{omega-log-divergence}
\ee
(Here $\theta (y)$ is the step function.)
If the soliton field, $U_c ({\bf x})$, is smooth in coordinate space,
then its Fourier transform, $\tilde{U}_c ({\bf k})$, decays
exponentially for large $|{\bf k}|$. This means that, as a function of
$|x|$, the integral in eq.(\ref{omega-log-divergence}) also decays
exponentially at large $|x|$.  One should keep in mind, however, that
the asymptotic formula, eq.(\ref{omega-log-divergence}), is valid only
in the parametric range $x \sim 1/N_c \ll \omega_0/M_N$ and thus does
not allow us to draw any conclusions about the behavior of the quark
distribution function at larger $|x|$.
\par
It is therefore interesting to derive the UV behavior of the
distribution function with energy cutoff at larger values of $|x|$.  An
asymptotic expansion can be performed in the domain 
$1/N_c \ll |x| \sim \omega_0/M_N$ by computing the UV divergences
of the moments of the
distribution function (see Appendix A). The asymptotic behavior is
given by
\be
[u(x) + d(x)]_{\rm occup.}^{\omega_0} &\sim&
N_c M_N^{-1} M^2 \frac{1}{24\pi^2}
\left[ \frac{4}{x} \delta\left( x + \frac{2\omega_0}{M_N}\right)
- \frac{1}{x^2} \theta\left( \frac{2\omega_0}{M_N} - |x| \right) 
\theta(-x)
\right]
\nonumber \\
&&
\times
\int d^3x \, \Tr\left[ \partial_k U ({\bf x}) 
\partial_k U^\dagger ({\bf x}) \right] , 
\nonumber \\
&& (1/N_c \ll |x| \sim \omega_0/M_N) .
\label{q-large-x-result}
\ee
It exhibits at large negative $x$ a rather slow $1/x^2$--decay, up to
the point $x = -2\omega_0/M_N$, where it ends with a
$\delta$--function peak. One should keep in mind that this
$\delta$--function appears only in the asymptotic limit 
$x \sim -\omega_0/M_N\to -\infty$. At large but finite values of
$x\sim -\omega_0/M_N$ the delta function in
eq.(\ref{q-large-x-result}) approximates a narrow peak whose width is
much smaller than $\omega_0/M_N$.  This ``large negative $x$''
behavior of the distribution function is an artifact of the energy
cutoff. We shall see below that both the $1/x^2$--tail and the delta
function cancel in Pauli-Villars regularization.
\par
In the previous section we have shown that the quark distribution
functions can be represented in two equivalent forms as a sum over
either occupied or non-occupied quark states.  In general,
regularization violates this equivalence.  Let us regularize the sum
over non-occupied states by an energy cutoff similar to
eq.(\ref{energy-cutoff-occupied}),
\be
[u(x) + d(x)]_{\rm non-occup.}^{\omega_0}
&=& -  N_c M_N \int\limits_{ E_{\rm lev}+0}^{\omega_0} d\omega
\Sp \left[ \delta(\omega - H) \delta(\omega - x M_N + p^3 )
(1 + \gamma^0 \gamma^3 ) \right]
\nonumber \\
&&
- (H\to H_0) .
\label{distribution-non-occupied-delta}
\ee
One can easily show that this sum over non-occupied states has the
same logarithmic divergence for large $\omega_0$ as the sum over
occupied states, eq.(\ref{omega-log-divergence}).  This means that the
difference between the two representations of the quark distribution
functions remains finite for $\omega_0\to \infty$.  The question is
whether this finite limit is zero or not.  The answer is,
surprisingly, no. Moreover, this limit can be computed analytically,
using a technique similar to the one described in Appendix A. One
finds
\be
\lefteqn{
\lim_{\omega_0\to\infty} \;
\left\{
[u(x) + d(x)]_{\rm occup.}^{\omega_0}
\, - \, [u(x) + d(x)]_{\rm non-occup.}^{\omega_0}
\right\} }
&& \nonumber\\
&=& N_c M_N M^2 \frac{1}{4\pi^2}  \int \frac{d^3k}{(2\pi)^3}
 \Tr \Bigl[
{\tilde U}({\bf k})  [{\tilde U}({\bf k})]^\dagger
\Bigr]
\log \frac{|x M_N + k^3 |}{|x M_N |} .
\label{omega-anomaly-result}
\ee
Thus, regularization by an energy cutoff leads to an anomalous difference 
between summation over occupied and non-occupied states even in 
the infinite--cutoff limit.
\par
The deeper reason for the artifacts encountered with the energy cutoff
is that this regularization violates the completeness of the set of
single--particle quark states. The equivalence
of the two representations of the distribution functions 
as sums over occupied and non-occupied states relies on the
locality of the equal--time anticommutator of quark fields (or,
equivalently, of the discontinuity of the Feynman Green function,
eq.(\ref{completeness})).  Leaving out the contribution of
high--energy states one is dealing with an incomplete set of quark
eigenstates, which results in a modification of the $\delta$--function
equal--time anticommutator. In other words, one violates 
causality, {\em i.e.}, the anticommutativity of the quark fields at
space--like separations. What is remarkable, though, is that cutoff 
regularization leads to anomaly--type phenomena which persist even in 
the infinite--cutoff limit, {\em cf.}\ eq.(\ref{omega-anomaly-result}). 
Furthermore, eq.(\ref{q-large-x-result}) tells us that such regularization
always leads to unphysical results for the distribution function, no matter
which representation of the distribution function one adopts.
It should be noted that the usual proper--time regularization of the
determinant is of this type.
\par
A regularization which preserves the completeness of states is the
Pauli--Villars regularization, where one subtracts from the divergent sums
a multiple of the corresponding sums over eigenstates of the Hamiltonian in 
which the quark mass, $M$, has been replaced by a regulator mass, $M_{PV}$. 
This mass now plays
the role of the physical cutoff of the effective theory, which was 
denoted generically by $\Lambda$ in Section 2. The coefficient of the
subtraction is chosen such as to cancel the logarithmic divergence of
the distribution function with the energy cutoff,
eq.(\ref{omega-log-divergence}).  We thus define
\be
&&
[u(x) + d(x)]^{PV}_{\rm occup.}(x)
\nonumber\\
&&
= \lim_{\omega_0\to\infty}
\left\{
[u(x)+d(x)]_{\rm occup.}^{\omega_0}(x) \,\, \rule[-1.0em]{.25mm}{2em}_{\, M}
- \frac{M^2}{M_{PV}^2} [u(x)+d(x)]_{\rm occup.}^{\omega_0}(x) \,\,
\rule[-1.0em]{.25mm}{2em}_{\, M_{PV}}
\right\}  ,
\label{PV}
\ee
and similarly for the sum over non-occupied states.  One observes that
the unphysical phenomena associated with the energy cutoff --- the
negative--$x$ behavior, eq.(\ref{q-large-x-result}), and the anomalous
difference of summing over occupied and non-occupied states,
eq.(\ref{omega-anomaly-result})) --- are proportional to the quark
mass squared, $M^2$. Thus, the artifacts of the energy cutoff
cancel under the Pauli--Villars subtraction, eq.(\ref{PV}), 
as it should be.  In
particular, the Pauli--Villars regularized distributions can now
equivalently be computed as sums over occupied or non-occupied states.
\par
In this section we have investigated the consequences of ultraviolet
regularization on the distribution functions using asymptotic
expansion techniques. The physical distributions, for finite
Pauli--Villars cutoff, can only be computed numerically. Below we
shall see that the numerically computed distribution functions in
Pauli--Villars regularization satisfy all general requirements. The
distributions decrease rapidly for large $|x|$ and exhibit the correct
positivity properties, in full accordance with the results of the
asymptotic analysis.
\section{Computation of quark distribution functions}
\setcounter{equation}{0}
\subsection{Spherically symmetric representation for distribution
functions}
We now develop a method for numerical computation of the
Pauli--Villars regularized distribution functions.  Our general
strategy will be as follows.  We compute the distribution functions as
sums over quark levels, eqs.(\ref{singlet_occ}, \ref{isovector_occ}),
for a large but finite energy cutoff. Such ``intermediate''
regularization is necessary in order to have expressions which can be
computed using finite basis methods. The physical distribution
functions are then obtained by subtracting the corresponding sums with
the PV regulator mass, $M_{PV}$, according to eq.(\ref{PV}), and
removing the energy cutoff by extrapolation to infinity.  In this way,
the energy cutoff affects only the intermediate steps of the
calculation, not the final result.
\par
For intermediate regularization we now introduce an energy cutoff in
Eqs.(\ref{singlet_occ}, \ref{isovector_occ}) in the form
\be
[u(x) + d(x)]_{\rm occup.}^R
&=& N_cM_N\!\!\sum\limits_{\scriptstyle n\atop \scriptstyle{\rm occup.}}\!
\langle n |  (1 + \gamma^0 \gamma^3 )
\delta( E_n - x M_N + p^3 ) | n \rangle
\, R ( E_n ) ,
\nonumber \\
\label{singlet_reg} \\
{[\Delta u(x) - \Delta d(x)]}_{\rm occup.}^R &=& -\frac{1}{3} (2T_3)  
N_c M_N \nonumber \\
&& \times
\!\!\sum\limits_{\scriptstyle n\atop \scriptstyle{\rm occup.}}\!
\langle n | \delta (E_n - x M_N + p^3 )
\tau^3 (1 + \gamma^0 \gamma^3 ) \gamma_5 | n \rangle
\, R ( E_n ) .
\nonumber \\
\label{isovector_reg}
\ee
We have written the matrix elements between single--particle levels in
abstract form, with $p^3$ denoting the $z$--component of the
single--particle momentum operator. Here, $R(E_n )$ is a smooth
regulator function with a cutoff, $E_{\rm max}$.  For example, one may
employ a Gaussian,
\be
R (E_n ) &=&
\exp \left( - \frac{E_n^2}{E_{\rm max}^2} \right) .
\label{gaussian}
\ee
Alternatively, one may use a Strutinsky (error function) regulator of
the kind described in ref.\cite{VJ90}, which leads to more rapidly
converging sums over levels.  A corresponding regularization can be
introduced also in the sums over non-occupied states.  
\par
Before evaluating the sums over levels, eqs.(\ref{singlet_reg},
\ref{isovector_reg}), it is convenient to convert them to a more
symmetric form. In the derivation of the distribution functions in
Section 2, using the infinite--momentum frame, it was assumed that the
nucleon is moving in the $z$--direction. The orientation of the
nucleon velocity is, of course, arbitrary, and the distribution
functions do not depend on it. We can thus write
eqs.(\ref{singlet_reg}, \ref{isovector_reg}) equivalently as
\be
[u(x) + d(x)]_{\rm occup.}^R
&=& N_cM_N\!\!\sum\limits_{\scriptstyle n\atop \scriptstyle{\rm occup.}}\!
\langle n | (1 + \gamma^0 \bfv\cdot\bfgamma )
\delta( E_n - x M_N + \bfv\cdot \bfp ) | n \rangle
\, R ( E_n ) ,
\nonumber \\
\label{singlet_v} \\
{[\Delta u(x) - \Delta d(x)]}_{\rm occup.}^R &=& -\frac{1}{3} (2T_3)  N_c M_N
\nonumber \\
&& \times
\!\!\sum\limits_{\scriptstyle n\atop \scriptstyle{\rm occup.}}\!
\langle n | \bfv\cdot\bftau (1 + \gamma^0 \bfv\cdot\bfgamma ) \gamma_5
\delta (E_n - x M_N + \bfp\cdot\bfv ) | n \rangle
\, R ( E_n ) .
\nonumber \\
\label{isovector_v}
\ee
where $\bfv$ is an arbitrary 3--dimensional unit vector, $\bfv^2 = 1$. 
For the isosinglet unpolarized distribution this is immediately obvious; in
the case of the isovector polarized distribution,
eq.(\ref{isovector_v}), we have made use of the ``hedgehog'' symmetry
of the classical meson field, eq.(\ref{hedge}), and the Hamiltonian,
eq.(\ref{H}), {\em i.e.}, the invariance under simultaneous rotations
in spin and isospin space.  We can now pass to a spherically symmetric
representation by averaging over the orientations of $\bfv$
computing the sum over quark levels. Using the identity
\be
\frac{1}{4\pi} \int d\Omega_v \delta ( E_n - x M_N + \bfv\cdot \bfp )
&=& \frac{1}{2 |\bfp |} \theta (|\bfp | - | E_n - x M_N |) ,
\ee
and its generalizations, we rewrite eqs.(\ref{singlet_v},
\ref{isovector_v}) in the form
\be
[u(x) + d(x)]_{\rm occup.}^R &=& N_c
M_N\!\!\sum\limits_{\scriptstyle n\atop \scriptstyle{\rm occup.}}\!
\langle n | A^{(1)} + A^{(2)} \gamma^0 \bfp\cdot\bfgamma | n \rangle
R ( E_n ) \, ,
\label{non_spherical}
\ee
\be
\lefteqn{ [\Delta u(x) - \Delta d(x)]_{\rm occup.}^R } && \nonumber \\
&=& -\frac{1}{3} (2T_3)  N_c M_N  
\sum\limits_{\scriptstyle n\atop \scriptstyle{\rm occup.}}\!
\langle n | A^{(3)} \gamma_5 \bfp\cdot\bftau 
+ A^{(4)} \gamma^0 \bfp\cdot\bfgamma
\gamma_5 \bfp\cdot\bftau + A^{(5)} \gamma^0 \bftau\cdot\bfgamma | n \rangle 
R ( E_n ) \, .
\nonumber \\
\label{pol_spherical}
\ee
Here, $A^{(k)} , (k = 1, \ldots 5)$, are scalar functions of the magnitude 
of the single--particle momentum operator, $|\bfp |$, as well as of the 
level energy, $E_n$, and $x$,
\be
A^{(k)} (|\bfp |, E_n, x)
&=& \frac{1}{2 |\bfp |} \theta (|\bfp | - | E_n - x M_N |)
\times
\left\{
\ba{lr}
1 , & k = 1 \\[2ex]
-{\displaystyle\frac{(E_n - x M_N)}{|\bfp |^2}} , & 2 \\[2ex]
-{\displaystyle\frac{(E_n - x M_N)}{|\bfp |^2 }}, & 3 \\[2ex]
-{\displaystyle\frac{1}{2 |\bfp |^2}}
+ {\displaystyle\frac{3}{2}} 
{\displaystyle\frac{(E_n - x M_N)^2}{|\bfp |^4}} , & 4 \\[2ex]
{\displaystyle\frac{1}{2}}
- {\displaystyle\frac{1}{2}} 
{\displaystyle\frac{(E_n - x M_N)^2}{|\bfp |^2}} , & 5
\ea
\right.
\nonumber \\
\ee
These operator functions of $|\bfp |$ are understood in the usual
sense, as functions of the eigenvalues in a basis where the operator
is diagonal.
\subsection{Evaluation in a discrete basis}
The distribution functions, eqs.(\ref{non_spherical},
\ref{pol_spherical}), are sums of diagonal matrix elements of functions
of single--particle operators between eigenstates of the Dirac
Hamiltonian in the background pion field, eq.(\ref{H}). To evaluate
them numerically we employ a basis of eigenfunctions of the free Dirac
Hamiltonian,
\be
H_0 \phi_i &=& E_i^{(0)} \phi_i ,
\hspace{1.5cm}
H_0 \;\; = \;\; -i\gamma^0 \gamma^k \partial_k + M \gamma^0 .
\label{eigen_H0}
\ee
The basis is made discrete by placing the soliton in a 3--dimensional
spherical box of finite radius, imposing the Kahana--Ripka boundary
conditions on the surface \cite{KR84}. The eigenvalues and
eigenfunctions of the full Hamiltonian, eq.(\ref{H}), are then
determined by numerical diagonalization in the discrete basis,
\be
\sum_j H_{ij} c_{nj} &=& E_n c_{ni},  \\
H_{ij} &=& \int_{\rm box} d^3 x \; \phi_i^\dagger ({\bf x} ) H 
\phi_j ({\bf x} ),  \\
\Phi_n ({\bf x} ) &=& \sum_i c_{ni} \phi_i ({\bf x} ) .
\ee
Since the operator $|\bfp |$ is a function of the free Hamiltonian,
\be
|\bfp | &=& \sqrt{H_0^2 - M^2},
\ee
it is diagonal in the basis of $H_0$ eigenstates, eq.(\ref{eigen_H0}), 
and one has
\be
\langle i | f(|\bfp |) | j \rangle &=& f(|\bfp |_i ) \delta_{ij},
\hspace{2cm} |\bfp |_i \;\; \equiv \;\; \sqrt{(E_i^{(0)})^2 - M^2} ,
\ee
for any function, $f(|\bfp |)$. Using this property one can
explicitly evaluate the matrix elements between levels in 
eqs.(\ref{non_spherical}, \ref{pol_spherical}), and obtains
\be
\lefteqn{[ u(x) + d(x) ]_{\rm occup.}^R} && \nonumber \\
&=& N_c
M_N\!\!\sum\limits_{\scriptstyle n\atop \scriptstyle{\rm occup.}}\!
\sum_{i, j} c_{ni}^{*} c_{nj}
\left[ A^{(1)} (|\bfp |_i , E_n , x) \delta_{ij}
+ A^{(2)} (|\bfp |_i , E_n, x) \, (\gamma^0 \bfp\cdot\bfgamma )_{ij} \right] 
R(E_n ) ,
\nonumber \\
\label{non_sum} \\
\lefteqn{ [\Delta u(x) - \Delta d(x) ]_{\rm occup.}^R } && \nonumber \\
&=& -\frac{1}{3} (2T_3)  N_c M_N
\!\!\sum\limits_{\scriptstyle n\atop \scriptstyle{\rm occup.}}\!
\sum_{i, j} c_{ni}^{*} c_{nj}
\left[ A^{(3)} (|\bfp |_i , E_n , x) (\gamma_5 \bfp\cdot\bftau )_{ij} \right. 
\nonumber \\
&+& \left.
A^{(4)} (|\bfp |_i , E_n , x) \sum_k (\gamma^0 \bfp\cdot\bfgamma )_{ik}
(\gamma_5 \bfp\cdot\bftau )_{kj}
+ A^{(5)} \left( |\bfp |_i , E_n , x \right) 
(\gamma^0 \bftau\cdot\bfgamma )_{ij}
\right] R(E_n ). 
\nonumber \\
\label{pol_sum}
\ee
Now the $A^{(k)} , (k = 1,\ldots 5)$ are ordinary functions of the
eigenvalues of $|\bfp |$ in basis states, $|\bfp |_i$.  Here, 
$(\ldots )_{ij}$ denote the matrix elements of the corresponding
operator between basis states. We remind that the corresponding
antiquark distributions are given, respectively, by minus the r.h.s.\
of eq.(\ref{non_sum}) at negative $x$, and that of eq.(\ref{pol_sum})
at negative $x$, see end of Section 2.2.
\par
We note that the Dirac and isospin structures appearing in
eqs.(\ref{non_sum}, \ref{pol_sum}) are essentially the same as those
in the sums determining the nucleon mass and isovector axial coupling,
$g_A^{(3)}$ \cite{DPP88,Review}. To simplify the calculation of the
matrix elements one may use that
\be
\gamma^0 \bfp\cdot\bfgamma &=& H_0 - M \gamma^0 ,
\ee
and that the operator appearing in the first and second term on the
r.h.s.\ of eq.(\ref{pol_sum}) can be expressed as an anticommutator,
\be
\gamma_5 \bfp\cdot\bftau &=& \frac{1}{2} \left\{ \gamma_5 \gamma^0
\bftau\cdot\bfgamma \, , \, \gamma^0 \bfp\cdot\bfgamma \right\} .
\ee
\par
The terms in the sums eqs.(\ref{non_sum}, \ref{pol_sum}) are proportional
to a step function depending on the level energy, $E_n$, the momenta of the
basis states, $|{\bf p}|_i$, and the Bjorken variable, $x$. The expressions
can not directly be used for numerical evaluation in a discrete basis,
since the result would be a discontinuous function of $x$.  There are,
however, ways to convert eqs.(\ref{non_sum}, \ref{pol_sum}) to a form
suitable for evaluation in a discrete basis. One possibility is to apply
Gaussian smearing in $x$ to the distribution functions.  Let us define
\be
D^{\rm smeared} (x) &\equiv& \frac{1}{\gamma\sqrt{\pi}}
\int_{-\infty}^{\infty} dx'\;
\exp \left(\frac{-(x - x')^2}{\gamma^2} \right)\; D(x') ,
\label{smearing}
\ee
where $D(x)$ stands for the regularized isosinglet unpolarized or 
isovector polarized 
distributions. Here, $\gamma \ll 1$ is a small but finite number.  These
``smeared'' distribution functions can now be calculated using 
eqs.(\ref{non_sum}, \ref{pol_sum}) with $A^{(k)}$ replaced by the corresponding
``smeared'' functions in $x$,
\beq
A^{(k)} (|\bfp |_i , E_n, x) \;\; \rightarrow \;\;
A^{(k) \, {\rm smeared}} (|\bfp |_i , E_n, x) .
\label{A_smearing}
\eeq
These are now continuous functions of the level momenta and energies, so
one may perform the sums over levels in the discrete basis, provided one
makes sure that the separation between the momentum eigenvalues of the
basis states is significantly smaller than the smearing width,
$\gamma$. The level spacing is inversely proportional to the size of the
Kahana--Ripka box, so it becomes necessary to use rather large boxes to
attain small values of $\gamma$. In the calculations described in
this paper we use a value of $\gamma = 0.1$, which requires box sizes $> 20
M^{-1}$.
\par
In this way one can compute the smeared distribution functions,
eq.(\ref{smearing}). At values of $x$ where the exact distributions are
smooth the smeared functions provide an excellent approximation to the
exact ones. An exception is the isosinglet unpolarized distribution near 
$x = 0$. The exact distribution has a discontinuity at $x = 0$, which
becomes a smooth crossover of width $\sim 1/\gamma$ in the smeared
distribution. It is possible to recover the discontinuity by
``deconvoluting'' the numerically computed smeared distributions. Dividing
the Fourier transform in $x$ of the numerically computed smeared
distribution by that of the Gaussian, eq.(\ref{smearing}), one can
reconstruct the Fourier transform of the exact distribution for values of
the argument up to $\sim 1/\gamma$. The exact distribution function
itself is then obtained by inverse Fourier transformation, incorporating
the known asymptotic behavior of the Fourier transform for large arguments
corresponding to a discontinuity at $x = 0$.
\par
We thus compute the mode sums for the smeared distributions,
eqs.(\ref{non_sum}, \ref{pol_sum}), for a number of values of the energy
cutoff, typically up to $E_{\rm max} \simeq 10\, M$, and also the
corresponding sums with the constituent quark mass replaced by the PV
regulator mass, $M_{PV}$. We then perform the PV subtraction,
eq.(\ref{PV}), and remove the energy cutoff by numerical extrapolation to
$E_{\rm max} \rightarrow \infty$ pointwise in $x$. One computes a
least--squares fit of the PV subtracted sums to a constant plus inverse
powers of $E_{\rm max}$, for each $x$.  The stability of the extrapolation
can be checked by adding more terms to the fit.
\par
In Section 3 we investigated the asymptotic behavior of the distribution
functions with an energy cutoff and noted a number of unphysical features,
which are removed by the Pauli--Villars subtraction. This can also be seen
directly in the numerical calculations. The numerically computed distribution 
functions for finite energy cutoff, eqs.(\ref{non_sum}, \ref{pol_sum}), 
exhibit a ``tail'' at large negative $x$, which is proportional 
to $M^2$, consistent with the asymptotic formula,
eq.(\ref{q-large-x-result}).  (For summation over non-occupied states, the
``tail'' occurs at positive $x$.) Moreover, the result for the anomalous
difference between summation over occupied and non-occupied states,
eq.(\ref{omega-anomaly-result}), is confirmed by numerical
calculations. Thus, the numerical results fully support the conclusions of
Section 3.
\par
Given the equivalence of summing over occupied and non-occupied states
in PV regularization, one may choose any of the two representations
for the numerical calculations. In practice, it is convenient to
compute the quark distributions by summing over occupied states and
the antiquark distributions by summing over non-occupied states. In
this way, no vacuum subtraction is required. Furthermore, these sums
exhibit asymptotic behavior in the energy cutoff earlier than the
respective other representations, making the extrapolation to infinite
cutoff more stable.
\par
When computing the isovector polarized distribution one must keep in mind
that it is defined as the limit of zero pion mass of the distribution
computed for finite pion mass, {\em i.e.}, for a soliton profile vanishing
exponentially at large radii \cite{DPPPW96}.  (The same limit is understood
in the definition of the isovector axial coupling, $g_A^{(3)}$.) We thus
must carry out the entire calculation described above (that is, summing
over quark levels, PV subtraction and extrapolation to 
$E_{\rm max} \rightarrow \infty$) for $M_\pi \neq 0$, and take the limit
$M_\pi \rightarrow 0$ by numerical extrapolation at the very
end\footnote{If one computed the isovector distribution directly for a
massless soliton profile, $P(r) \sim 1/r^2$ for $r \rightarrow \infty$, one
would find a singularity at $x = 0$ (regulated only by the finite box
size). The distribution obtained as a limit of a massive profile is
non-singular at $x = 0$.}.
\par
The Gaussian smearing, eq.(\ref{smearing}), offers a simple possibility to
compute the distribution function directly as a function of $x$. We note,
however, that the use of the spherically symmetric representation,
eqs.(\ref{non_sum}, \ref{pol_sum}), is not limited to this method. In fact,
performing other functional transformations of the expressions
eqs.(\ref{non_sum}, \ref{pol_sum}) before summing over levels one can
obtain prescriptions for evaluating the distribution functions in a variety
of representations. For example, replacing in eqs.(\ref{non_sum},
\ref{pol_sum}) the functions $A^{(k)}$ by their moments,
\beq
A^{(k)}_m (|\bfp |_i , E_n) \;\; = \;\;
\int_{-1}^1 dx x^{m-1} A^{(k)} (|\bfp |_i , E_n, x) , 
\hspace{1cm} (m = 1, 2, \ldots ),
\label{A_moments}
\eeq
one obtains a formula for numerical evaluation of the moments of the
distribution function. (Again, one must compute
the sums for finite energy cutoff, perform the Pauli--Villars subtraction
and extrapolate to infinite cutoff.) We have computed the lowest moments
of the distributions in this way $(m < 10)$ and verified that they coincide
with the moments of the numerically computed distribution functions.
\section{Numerical results and discussion}
\setcounter{equation}{0}
In the numerical calculations we use the standard value for the constituent
quark mass, $M = 350 \, {\rm MeV}$, as derived from the instanton vacuum
\cite{DP86}. The value of the PV regulator mass, $M_{PV}$, is determined by
reproducing the experimental value of the pion decay constant,
\be
F_\pi^2 &=& 4N_c\int\frac{d^4k}{(2\pi)^4}\;\frac{M^2}{(M^2 + k^2)^2}
-4 N_c\frac{M^2}{M_{PV}^2}\int\frac{d^4k}{(2\pi )^4}\;
\frac{M_{PV}^2}{(M_{PV}^2 + k^2)^2} \nonumber \\
&=& \frac{N_c M^2}{4\pi^2} \log \frac{M_{PV}^2}{M^2} .
\label{fpi}
\ee
With $F_\pi = 93 \, {\rm MeV}$ one obtains $M_{PV}^2 /M^2 = 2.52$.  For the
soliton profile, eq.(\ref{hedge}), we use the variational form of
ref.\cite{DPP88},
\be
P(r) &=& -2\;\arctan\left(\frac{r_0^2}{r^2}\right) ,
\label{var}
\ee
with $r_0 = 1.0\, M^{-1}$, which gives a reasonable description of a
varitey of hadronic observables of the nucleon.  For these
parameters, the nucleon mass is found to be $M_N = 1150 \, {\rm MeV}$. (The
nucleon mass is also computed in PV regularization, subtracting from
eq.(\ref{M_N}) $M^2/M_{PV}^2$ times the corresponding expression for the
Hamiltonian with $M_{PV}$. The contribution of the discrete level is also
subtracted.)  For calculation of the isovector polarized distribution, we
introduce a finite pion mass in eq.(\ref{var}) in the form
\be
P_{M_\pi} (r) &=&
-2\;\arctan\left[ \frac{r_0^2}{r^2} (1 + M_\pi r ) \exp (-M_\pi r)
\right] .
\label{var_mpi}
\ee
This form has the correct Yukawa tail at large $r$ but is not modified
compared to eq.(\ref{var}) at $r = 0$. The limit $M_\pi \rightarrow 0$ is
taken at the very end of the calculation.
\par
The result for the isosinglet unpolarized quark-- and antiquark
distributions is shown in Fig.1. For both distributions we show 
separately the total result (the sum of the
discrete level and the negative Dirac continuum) and the contribution
of the discrete level.  One sees that the discrete level contributes
to the antiquark distribution with a negative sign.  (The contribution
of the discrete level to the r.h.s.\ of eq.(\ref{non_sum}) is
continuous at $x = 0$, and the antiquark distribution is just given by
the negative of eq.(\ref{non_sum}) at negative $x$.) An approximation
in which only the discrete level is taken into
account would thus lead to negative antiquark distributions
\cite{WGR}.  Positivity of the antiquark distribution is naturally
restored by including the Dirac continuum.  This
is clear in the light of the discussion of Section 3: Restricting
oneself to the contribution of the discrete level one is working with
an incomplete set of states. Only the sum of all levels (discrete plus
Dirac continuum) gives the correct realization of the distribution
function in the effective theory.
\par
The result for the isovector polarized quark-- and antiquark
distributions is displayed in Fig.2 (total results and level
contributions). One again observes a sizable contribution from the
Dirac continuum, which reverses the sign of the level contribution to
the antiquark distribution.  Here, however, contrary to the isosinglet
unpolarized distribution, no definite sign is required a priori.
\par
The calculated distributions should in principle be used as input for
perturbative evolution, starting with a scale of the order of the cutoff,
$M_{PV} \simeq 600 \, {\rm MeV}$.  We stress that we are computing the
twist--2 parton distributions at a low normalization point, not the
structure function (cross section) at low $q^2$, so a meaningful comparison
with the data can be performed only after evolution to large
$q^2$. Alternatively, we may compare our calculations with the
parametrizations of Gl\"uck, Reya {\em et al.}
\cite{GRV95,GRSV96}. Starting from ``valence--like'' (non-singular)
quark--, antiquark and gluon distributions at a normalization point well
below $1 \, {\rm GeV}$, these authors can fit at large $q^2$ not only all
the data in the large--$x$ region, but also the recent small--$x$ data down
to $x \sim 10^{-4}$.  We emphasize that the quark-- and antiquark
distributions obtained in our approach are
precisely of this ``valence--like'' form. Moreover, the normalization
points of the LO and NLO distributions of \cite{GRV95,GRSV96} are close to
our cutoff, so one may perform a preliminary comparison without taking into
account evolution.
\par
Fig.3 shows the isosinglet unpolarized total distribution (quarks plus
antiquarks) together with the fits of \cite{GRV95}.  Our distribution is
larger than that of \cite{GRV95} since their fit includes gluons, which
carry about 30 percent of the nucleon momentum at this scale. For
the variational soliton profile, eq.(\ref{var}), the second moment of the
calculated distribution of quarks plus antiquarks is $0.8$. (With a
self--consistent solution it would be unity, since the energy momentum sum
rule follows from the equations of motion for the pion field
\cite{DPPPW96}.).
\par
The isosinglet unpolarized valence quark distribution (quarks minus
antiquarks) is compared in Fig.4. Here we have taken in our calculation
$M_{PV} \rightarrow \infty$, since this distribution function is
ultraviolet finite and should not be regularized in order to preserve the
baryon number sum rule.  It is interesting to see how this sum rule is
realized in the large--$N_c$ approach. In fact, the (unregularized)
contribution of each level to the baryon number is $N_c$, as can be seen,
for instance, by integrating eq.(\ref{non_spherical}) over both negative
and positive $x$.  Since the baryon number of the discrete level is $N_c$,
the contribution of the negative Dirac continuum to the baryon number must
be zero in order to satisfy the sum rule for the total
distribution. Indeed, one observes that the baryon number of the negative
continuum vanishes when the cutoff is taken to infinity. However,
this does not mean that the Dirac continuum does not contribute to the
valence quark distribution --- just the integral of its contribution is
zero. In Fig.4 we show both the total result (discrete level plus
continuum) and the continuum contribution, which integrates to zero.
\par
The isovector polarized total distribution (quarks plus antiquarks) is
shown in Fig.5. The calculated distribution is systematically smaller
than the fit of \cite{GRSV96}. This is related to the fact that the
isovector axial coupling, $g_A^{(3)}$, which determines the
normalization of this distribution, is underestimated in the leading
order of the $1/N_c$--expansion (with our parameters for the chiral
soliton we obtain $g_A^{(3)} = 0.9$). We note that $1/N_c$ corrections
to this quantity have been computed \cite{WW93}; the same techniques
could also be applied to the distribution functions. In Fig.6 we show
the polarized antiquark distribution, which was assumed to be zero in
the fit of \cite{GRSV96}. In our calculation it is obtained non-zero,
however, significantly smaller than the total distribution of quarks
plus antiquarks.
\par
To summarize, we obtain a reasonable description of the isosinglet
unpolarized and isovector polarized quark and antiquark distributions.  In
particular, we find a large antiquark distribution at the low normalization
point, in agreement with the parametrizations of the data.
\par
In the calculations reported here we have chosen the variational soliton
profile, eq.(\ref{var}), with a radius which gives a reasonable overall
description of a number of hadronic observables, for example the
$N\Delta$--splitting \cite{DPP88}.  In principle, the classical pion field
describing the nucleon should be determined as the minimum of the static
energy, {\em i.e.}, as the self--consistent solution of the equations of
motion of the pion field.  The calculation of parton distributions with the
self--consistent pion field will be the subject of a separate
investigation.
\par
The exact numerical calculations fully support the approximation used in
\cite{DPPPW96}, based on the ``interpolation formula'' for the quark
propagator in the background pion field.  For both unpolarized and polarized
distributions the differences to the exact results are of the order of 10
percent for the total distributions (quarks plus antiquark), somewhat
larger for quarks and antiquarks separately.
\section{Conclusions}
\setcounter{equation}{0}
In this paper we have completed the first part of the program formulated
in \cite{DPPPW96}. We have computed the leading quark distribution
functions in the $1/N_c$--expansion, namely the isosinglet unpolarized
and isovector polarized, in the effective chiral theory.
\par
Starting from the original definition of parton distributions as
numbers of particles fraction $x$ of the nucleon momentum in the
infinite--momentum frame, we have shown that it leads in the
large--$N_c$ limit to the same results as the QCD definition of
distribution functions as matrix elements of bilinears of quark fields
on the light cone.  The fact that the equivalence of the two
definitions of distribution functions can be established within the
mean--field picture at large $N_c$ shows the scope of this
relativistically covariant, field--theoretical description of the
nucleon.
\par
We have observed that, generally speaking, the calculation of quark
distribution functions puts strong demands on the regularization of the
effective theory. A crucial requirement is that it should preserve the
completeness of the basis of single--particle quark wave functions in which
one expands the fermion fields in the mean--field approximation.  A
regularization by subtraction, such as the Pauli--Villars regularization,
meets this requirement, while methods based on an energy (or other) cutoff
violate this completeness, and thus, causality. In particular, 
the anomaly observed in the difference of summation over occupied and
non-occupied states in Section 3 shows that one faces
here a truly qualitative difference between regularization methods, not
simply finite--cutoff effects vanishing in the infinite--cutoff limit. By
explicit calculation, we have shown that Pauli--Villars regularization
leads to quark and antiquark distributions satisfying all general
requirements.
\par
As to the numerical calculations, we have presented a general scheme for
computing the distribution functions in various representations. It is
remarkable that the Kahana--Ripka method, using a basis of
eigenstates of the free Hamiltonian, lends itself so naturally to 
the computation of distribution functions after one has converted them 
to a spherically symmetric form.
\par
The methods developed here, both analytical and numerical, can readily be
generalized to compute also the ``small'' quark distributions in the
$1/N_c$--expansion, the isovector unpolarized and isosinglet polarized
distributions. They are given by double sums over quark levels in the
background pion field (formulas have been presented in \cite{DPPPW96}).
Calculations of these distributions are in progress.
\par
We have found reasonable agreement of our results with the fits of
Gl\"uck, Reya {\em et al.} \cite{GRV95,GRSV96}.  Indeed, the
large--$N_c$ approach to parton distributions formulated in
\cite{DPPPW96} provides justification for the picture of
``valence--like'' distributions at a low normalization point. The fact
that at large $N_c$ the nucleon is characterized by a classical pion
field (or, equivalently, a polarized Dirac sea of quarks) naturally
explains the large antiquark content at the low normalization
point. The antiquark distributions obtained in our approach
are non-singular at small $x$.  We note also that the parametric
suppression of the gluon relative to the quark distributions, which is
implied by the effective chiral theory (see Section 1 and
ref.\cite{DPPPW96}), seems not to be in contradiction with the
parametrization of the data at low normalization point \cite{GRV95}. A
30 percent momentum fraction of gluons at the low
normalization point is consistent with the suppression of the gluon
distribution by $M^2 / \Lambda^2$.  However, in order to make this
more quantitative one should develop this approach to a level which
allows one to compute a non-zero gluon distribution. This can be done
in the framework of the instanton vacuum, using the methods of
\cite{DPW96}.
\\[1.5cm]
{\large\bf Acknowledgements} \\[.3cm]
This work has been supported in part by the NATO Scientific Exchange grant
OIUR.LG 951035, by INTAS grants 93-0283 EXT and 93-1630-EXT, by a joint
grant of the Deutsche Forschungsgemeinschaft and the Russian Foundation for
Basic Research, and by COSY (J\"ulich). The Russian participants
acknowledge the hospitality of Bochum University.  P.V.P.\ and M.V.P.\ are
supported by the A.v.Humboldt Foundation.  It is a pleasure to thank Klaus
Goeke for encouragement and multiple help.
\appendix
\section{Asymptotics for large energy cutoff}
\renewcommand{\theequation}{\Alph{section}.\arabic{equation}}
\setcounter{equation}{0}
In this appendix we discuss the asymptotic properties of the
distribution function with the energy cutoff, which was introduced
in Section 3 as a device to control the intermediate steps of the
calculation. In particular, we show how the large--$x$ asymptotic behavior,
eq.(\ref{q-large-x-result}), can be derived.  This formula assumes the
double limit of large $x$ (in the sense of $N_c x\gg 1$) and large
energy cutoff ($\omega_0\gg M$). It is convenient to analyze this
limit in terms of the moments of the distribution functions.  An
explicit expression for the moments is obtained by integrating
eq.(\ref{energy-cutoff-occupied}) over $x$,
\be
M_n^{\omega_0}
&=& \int\limits_{-1}^{1}dx\, x^{n-1} [u(x)+d(x)]_{\rm occup.}^{\omega_0}
\nonumber \\
&=&  N_c M_N^{1-n}
\int\limits_{-\omega_0}^{ E_{\rm lev}+0} d\omega
\Sp\left[ \delta(\omega-H) (\omega + p^3)^{n-1}
(1+\gamma^0\gamma^3) \right]
- (H\to H_0), 
\label{Moment-occupied-again-3} 
\nonumber
\\
&& (n = 1, 2, \ldots ).
\ee
The delta function of the Hamiltonian can be represented as the
discontinuity of the quark propagator, {\em cf.}\
eq.(\ref{delta_to_propagator}). One obtains a representation of the
moments as
\beq
M_n^{\omega_0}
= - \, \frac{i}{2\pi}
\int\limits_{0}^{\omega_0} d\omega'
[{\tilde M}_n(-i\omega' + 0)
- {\tilde M}_n(-i\omega' - 0)] ,
\label{M-Im:tilde-M} 
\eeq
where
\be
{\tilde M}_n(\omega)  &=& -i N_c M_N^{1-n}
\Sp \left[
\frac{1}{-\omega+iH} (-i\omega + p^3)^{n-1}
(1+\gamma^0\gamma^3) \right]
- (H\to H_0).
\label{tilde-M-definition}
\ee
\par
The integrand of eq.(\ref{Moment-occupied-again-3}) is ultraviolet
finite for any fixed $\omega$, and an ultraviolet divergence appears
in eq.(\ref{Moment-occupied-again-3}) only in the limit $\omega_0
\rightarrow \infty$ due the large--$\omega$ behavior of the integrand
(typically a power--like growth, see
below). Eq.(\ref{tilde-M-definition}), on the other hand, contains
ultraviolet divergences even for fixed $\omega$ due to large momenta.
The difference is that in eq.(\ref{Moment-occupied-again-3}) we have
the delta function of the Hamiltonian, which constrains the
ultraviolet growth of the integrand. We therefore have to introduce
an additional regularization of the functional trace in
eq.(\ref{tilde-M-definition}) at fixed $\omega$. The dependence on
this additional cutoff cancels, however, after taking the
discontinuity in eq.(\ref{M-Im:tilde-M}), and thus does not influence
the final result.
\par
Writing the propagator as in eq.(\ref{propagator_squared}) one can
expand eq.(\ref{tilde-M-definition}) in derivatives of the pion field,
\be
{\tilde M}_n(\omega)
&=& \sum\limits_{k=1}^\infty
{\tilde M}_n^{(k)}(\omega)
\label{Moment-O4-5-A-reg} , \\
{\tilde M}_n^{(k)}(\omega)
&=& (-i)^n N_c M_N^{1-n}
\Sp \left\{ \frac{1}{\omega^2- \partial_k^2 + M^2}
\left[ M(\partialslash U^{\gamma_5})
\frac{1}{\omega^2- \partial_k^2 + M^2} \right]^k \right.
\nonumber \\
&&
\times \left.
(- i\omega\gamma_0 - \gamma_k\partial_k + iMU^{-\gamma_5})
(\omega + \partial_3 )^{n-1}
(-i\omega\gamma_0 + \gamma_3 ) \right\} .
\label{Moment-O4-5-B-reg}
\ee
For the calculation of the leading large--$x$ asymptotics it is
sufficient to restrict oneself to the first two terms in
eq.(\ref{Moment-O4-5-A-reg}). The traces are easily evaluated by
inserting plane--wave states,
\be
{\tilde M}_n^{(k)}(\omega )
&=& 4 M^2 N_c M_N^{1-n}
\int \frac{d^3p}{(2\pi)^3}
\int\limits \frac{d^3q}{(2\pi)^3} 
\Tr \left[ \tilde U({\bf p}) [\tilde U({\bf p})]^\dagger \right]
\frac{1}{[\omega^2+({\bf p}-{\bf q})^2+M^2]}
\nonumber \\
&& \times \left\{ 
\ba{lr}
{\displaystyle\frac{p^3 (-i\omega + q^3 )^{n-1}}
{(\omega^2 + |{\bf q}|^2 + M^2)}}, & k = 1, \\[2ex]
{\displaystyle\frac{(-i\omega + q^3 )^n |{\bf p}|^2}
{(\omega^2 + |{\bf q}|^2 + M^2)^2}}, & k = 2.
\ea \right.
\ee
The integral over $q$ is divergent, but, as said above, the divergent
terms do not contribute to the discontinuity in $\omega$. Keeping only
those terms that lead to the discontinuity we obtain
\be
{\tilde M}_{n}^{(k)}(\omega)
&=& 4 M^2 N_c M_N^{1-n}
(-i)^{n+2}
\omega^{n-3}
\mbox{sign}(\mbox{Re}\,\omega)
\int d^3x \Tr\left[ \partial_k U ({\bf x}) 
\partial_k U^\dagger ({\bf x}) \right]
\nonumber \\
&&
\times
\left\{ 
\ba{lr}
{\displaystyle\frac{1}{32\pi}} 
\left[ 2^{n-2} (n-2) - {\displaystyle\frac{1}{2}} \delta_{n,1} \right] ,
& k = 1, \\[2ex]
{\displaystyle\frac{1}{48\pi}}
\left[  2^{n-3}  (n-1) + {\displaystyle\frac{1}{2}} \delta_{n,2} \right] ,
& k = 2.
\ea \right.
\ee
Inserting this asymptotic behavior in eq.(\ref{M-Im:tilde-M}) we
obtain the leading divergences of the moments for 
$\omega_0 \rightarrow \infty$,
\be
M_{n}^{\omega_0}
&\sim&
M^2 N_c M_N^{-1} \int d^3x 
\Tr\left[ \partial_k U ({\bf x}) \partial_k U^\dagger ({\bf x} ) \right]
\nonumber  
\\
&& \times \left\{ \ba{lr}
(-1)^n {\displaystyle\frac{1}{48\pi^2}} \left(
{\displaystyle\frac{2\omega_0}{M_N}} \right)^{n-2}
\left[ 8 + {\displaystyle\frac{2}{n-2}} \right], & n \ge 3, \\[2ex]
{\displaystyle\frac{1}{12\pi^2}} \log\omega_0 , & n = 2.
\ea \right.
\label{moments_div}
\ee
Thus, the moments for $n \ge 3$ have power divergences with
the energy cutoff, while for $n = 2$ the divergence is logarithmic.
\par
The asymptotic behavior of the moments may be expressed in the form 
of a function of $x$, assuming that $1/N_c \ll |x| \sim \omega_0/M_N$.
Computing the moments of the function of eq.(\ref{q-large-x-result})
one may easily check that it corresponds to the large--$\omega_0$
behavior of the $n > 3$ moments, eq.(\ref{moments_div}).
\par
To conclude, we have shown that, with an energy cutoff, the moments of
the distribution function generally have power divergences. These
manifest themselves not in a power divergence of the distribution
function at fixed $x$, but in the occurrence of a ``tail'' at 
large negative $x$ (large positive $x$ in the case of summation over
non-occupied states).  The power divergences are artifacts of the
energy cutoff and cancel under the Pauli--Villars subtraction,
eq.(\ref{PV}).  The physical, Pauli--Villars regularized distribution
functions are well--localized in $x$.
%

%
%
\newpage
\begin{figure}
\vspace{-1cm}
\epsfxsize=16cm
\epsfysize=15cm
\centerline{\epsffile{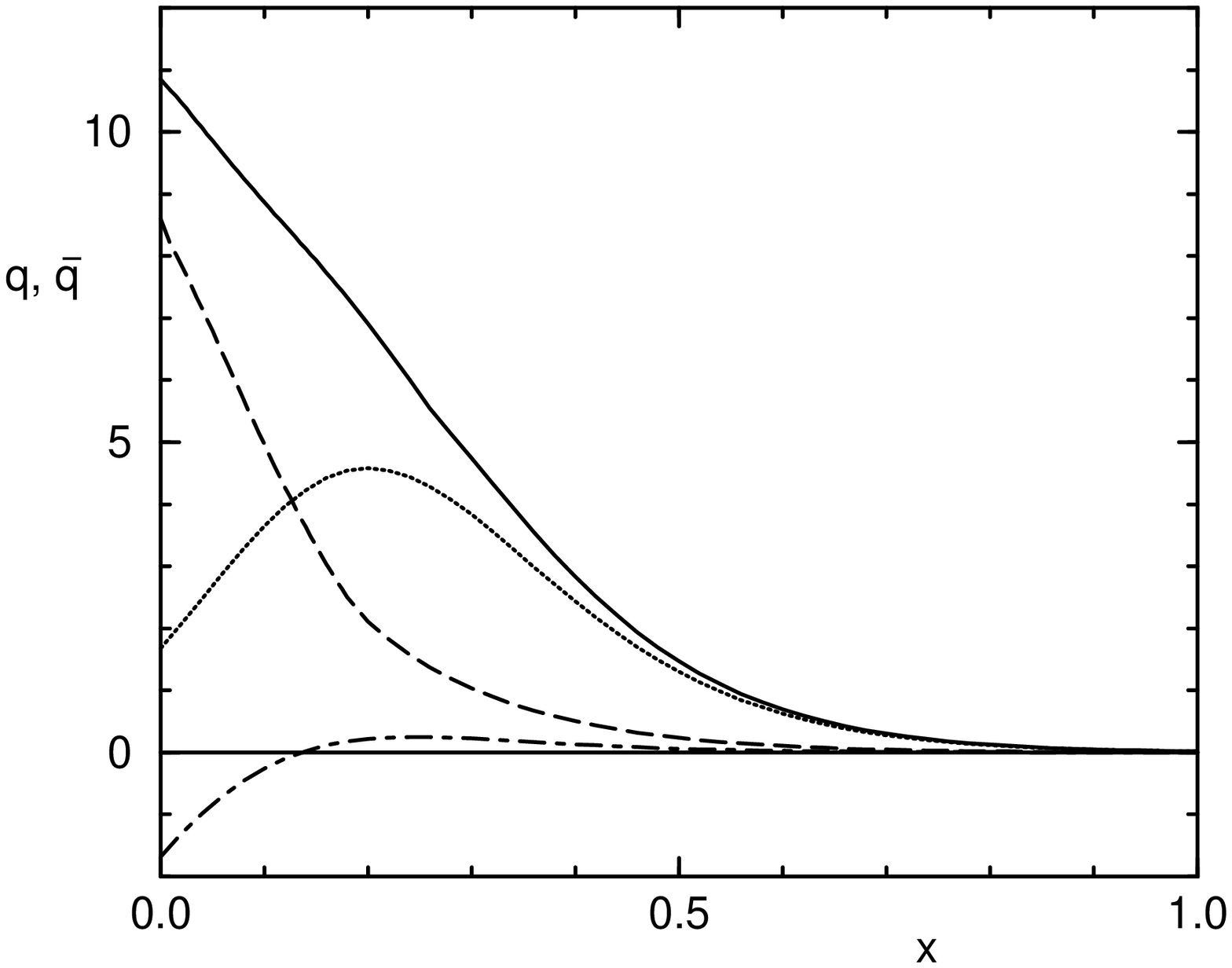}}
\caption[Isosinglet unpolarized distribution, full function]
{The isosinglet unpolarized quark-- and antiquark distributions.  
{\em Solid line:} quark distribution, $u(x) + d(x)$, total result
(discrete level plus Dirac continuum); 
{\em dotted line:} contribution of the discrete level (after PV 
subtraction) to $u(x) + d(x)$.  
{\em Dashed line:} antiquark distribution, $\bar u(x) + \bar d(x)$, 
total result; 
{\em dot--dashed line:} contribution of the discrete level to 
$\bar u(x) + \bar d(x)$.}
\label{fig_non}
\end{figure}
\newpage
\begin{figure}
\vspace{-1cm}
\epsfxsize=16cm
\epsfysize=15cm
\centerline{\epsffile{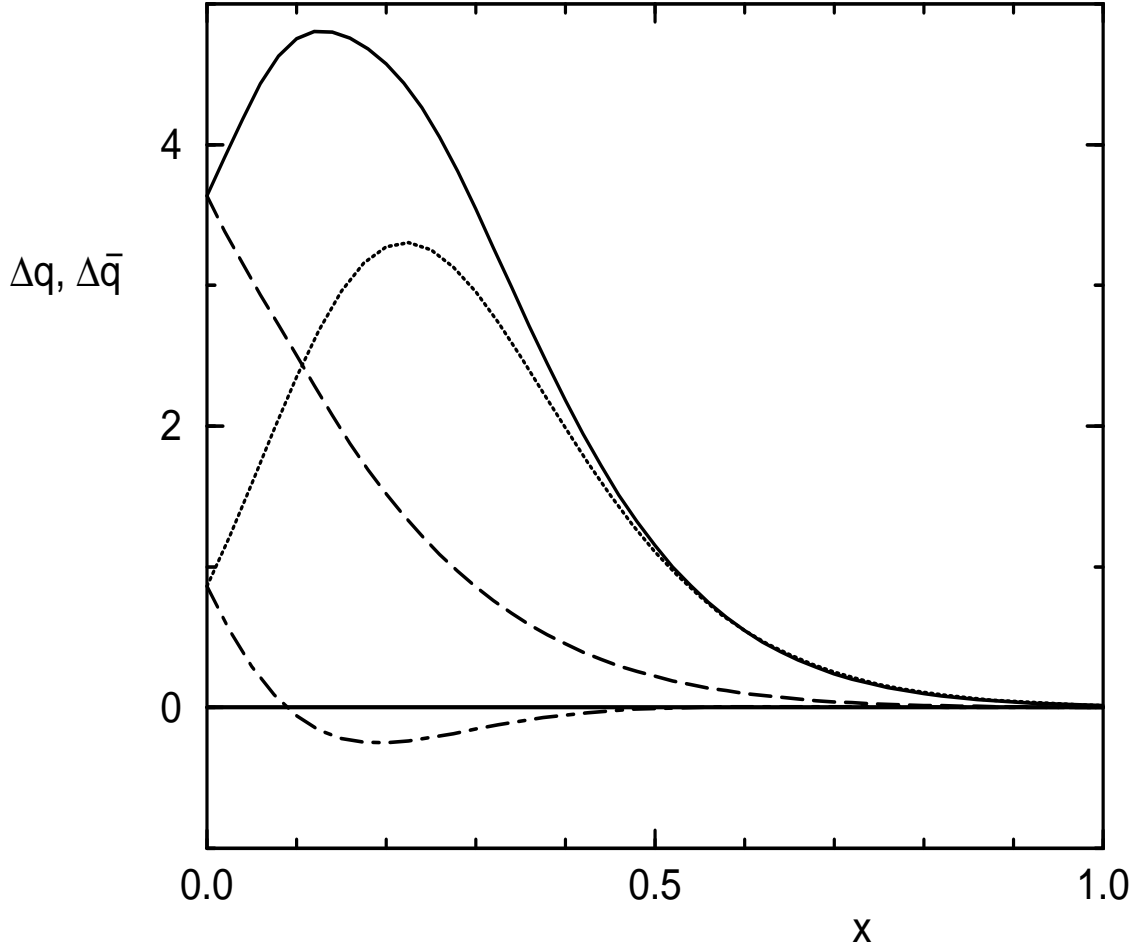}}
\caption[Isovector polarized distribution, full function]
{The isovector polarized quark-- and antiquark distributions.
{\em Solid line:} quark distribution, $\Delta u(x) - \Delta d(x)$, total 
result (discrete level plus Dirac continuum);
{\em dotted line:} contribution of the discrete level 
(after PV subtraction) to $\Delta u(x) - \Delta d(x)$.
{\em Dashed line:} antiquark distribution, 
$\Delta\bar u(x) - \Delta\bar d(x)$, total result;
{\em dot--dashed line:} contribution of the discrete level to 
$\Delta\bar u(x) - \Delta\bar d(x)$.}
\label{fig_pol}
\end{figure}
\newpage
\begin{figure}
\vspace{-1cm}
\epsfxsize=16cm
\epsfysize=15cm
\centerline{\epsffile{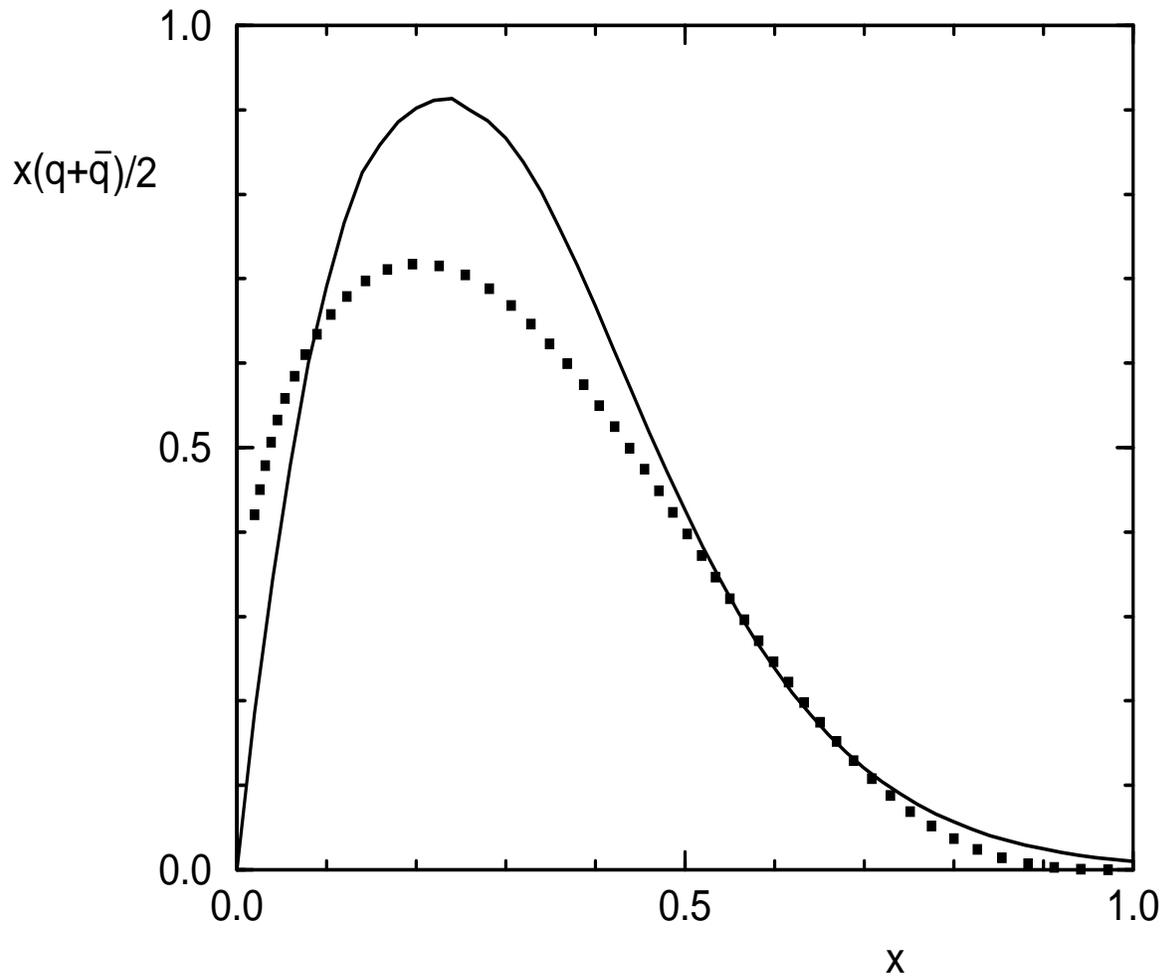}}
\caption[Isosinglet unpolarized distribution,
quarks plus antiquarks]
{The isosinglet unpolarized distribution of quarks plus antiquarks,
$\frac{1}{2} x [u(x) + d(x) + \bar{u} (x) + \bar{d} (x)]$.
{\em Solid line:} calculated distribution (total result, 
{\em cf.}\ Fig.\ref{fig_non}).
{\em Points:} NLO parametrization of ref.\cite{GRV95}.}
\end{figure}
\newpage
\begin{figure}
\vspace{-1cm}
\epsfxsize=16cm
\epsfysize=15cm
\centerline{\epsffile{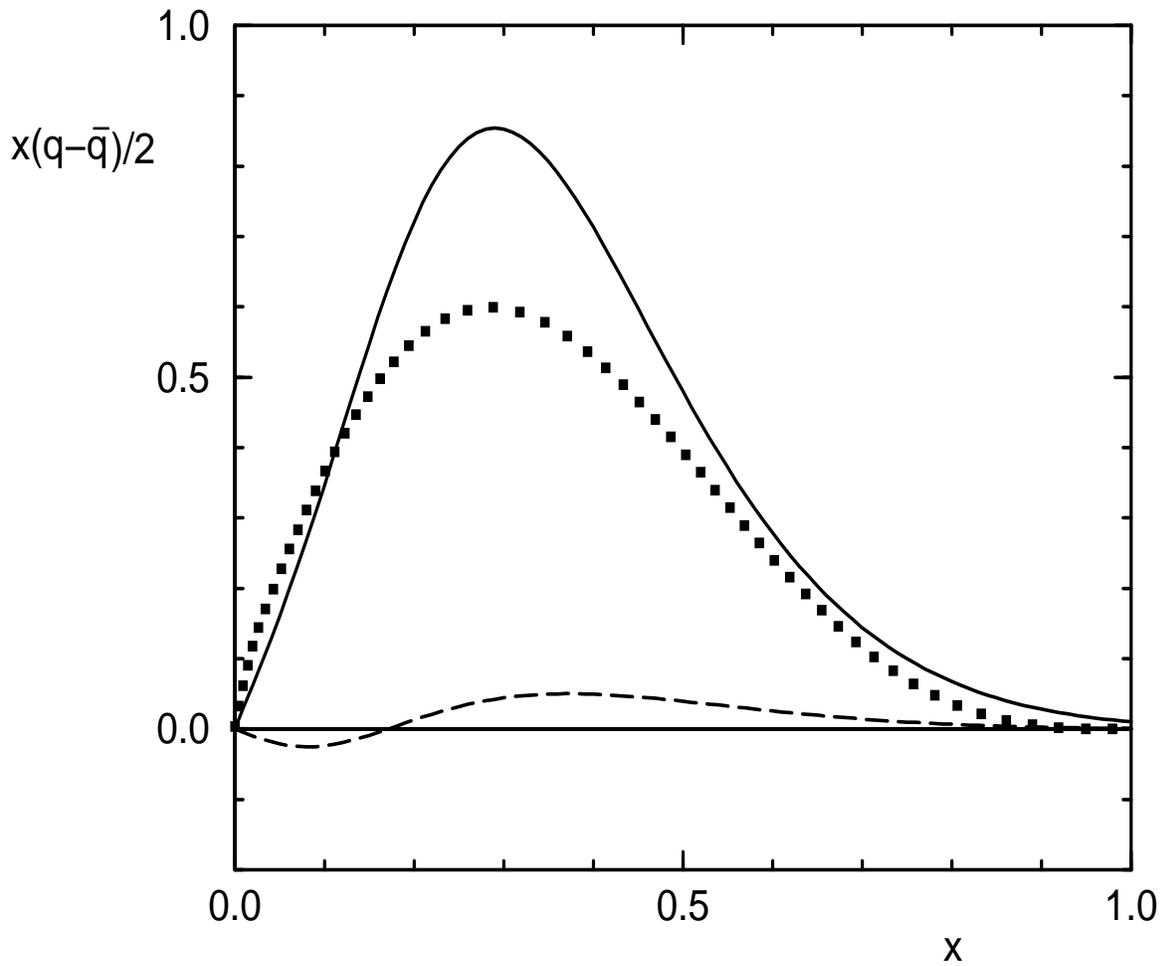}}
\caption[Isosinglet unpolarized distribution, valence]
{The isosinglet unpolarized valence quark distribution,
$\frac{1}{2} x [u(x) + d(x) - \bar{u} (x) - \bar{d} (x)]$.
{\em Solid line:} calculated distribution (total result);
{\em dashed line:} contribution of the Dirac continuum.
{\em Points:} NLO parametrization of ref.\cite{GRV95}.}
\end{figure}
\newpage
\begin{figure}
\vspace{-1cm}
\epsfxsize=16cm
\epsfysize=15cm
\centerline{\epsffile{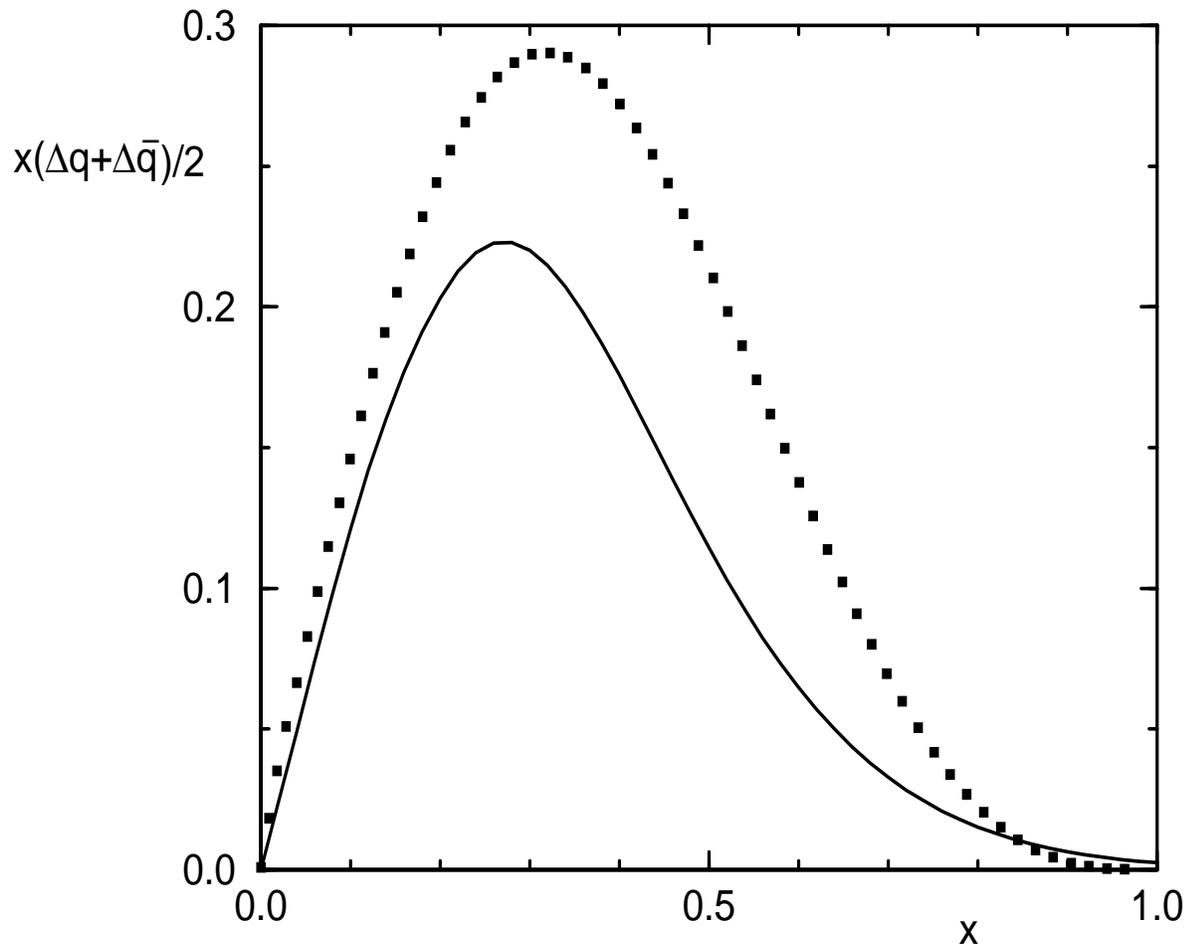}}
\caption[Isovector polarized distribution, quarks plus antiquarks]
{The isovector polarized distribution of quarks plus antiquarks,
$\frac{1}{2} x [\Delta u(x) - \Delta d(x) + \Delta \bar{u} (x)
- \Delta \bar{d} (x)]$.
{\em Solid line:} calculated distribution (total result, 
{\em cf.}\ Fig.\ref{fig_pol}).
{\em Points:} NLO parametrization of ref.\cite{GRSV96}.}
\end{figure}
\newpage
\begin{figure}
\vspace{-1cm}
\epsfxsize=16cm
\epsfysize=15cm
\centerline{\epsffile{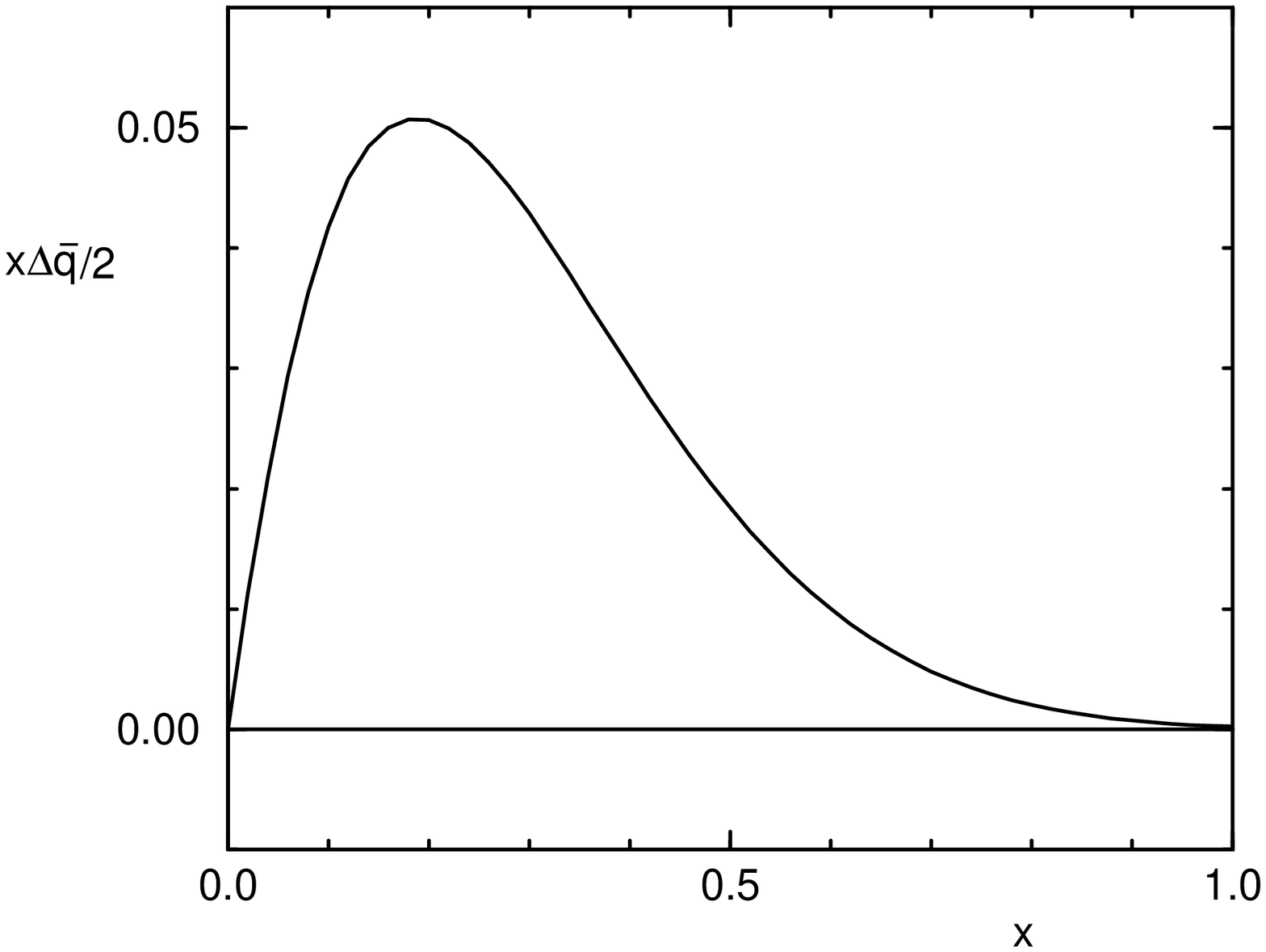}}
\caption[Isovector polarized distribution, antiquarks]
{The isovector polarized antiquark distribution,
$\frac{1}{2} x [\Delta \bar u(x) - \Delta \bar d(x)]$.
{\em Solid line:} calculated distribution (total result, 
{\em cf.}\ Fig.\ref{fig_pol}). In the fit of ref.\cite{GRSV96} 
this distribution is assumed to be zero.}
\end{figure}

\newpage
\begin{thebibliography}{99}
%
\bibitem{MRS95}
For a review, see: A.D.\ Martin, Acta Phys.\ Pol.\ {\bf 27} (1996) 1287; \\
A.D.\ Martin, R.G.\ Roberts, and W.J.\ Stirling, 
Phys.\ Lett.\ {\bf B 354} (1995) 155; Phys.\ Rev.\ {\bf D 50} (1994) 6734.
%
\bibitem{CTEQ95}
The CTEQ collaboration: H.L.\ Lai {\em et al.}, Phys.\ Rev.\ {\bf D 55}
(1997) 1280; Phys.\ Rev.\ {\bf D 51} (1995) 4763.
%
\bibitem{GRV95}
M. Gl\"uck, E. Reya, and A. Vogt, Z. Phys.\ {\bf C 67} (1995) 433.
%
\bibitem{GRSV96}
M. Gl\"uck, E. Reya, M. Stratmann, and W. Vogelsang, Phys.\ Rev.\
{\bf D 53} (1996) 4775.
%
\bibitem{DPPPW96} D.I.\ Diakonov, V.Yu.\ Petrov, P.V.\ Pobylitsa,
M.V.\ Polyakov, and C. Weiss, Nucl.\ Phys.\ {\bf B 480} (1996) 341.
%
\bibitem{DPP88}
D.I.\ Diakonov, V.Yu.\ Petrov, and P.V.\ Pobylitsa,
Nucl.\ Phys.\ {\bf B 306} (1988) 809.
%
\bibitem{Review}For a review, see: Ch.V.\ Christov {\em et al.},
Prog.\ Part.\ Nucl.\ Phys.\ {\bf 37} (1996) 91.
%
\bibitem{DPW96}
D. Diakonov, M. Polyakov, and C. Weiss, Nucl.\ Phys.\ {\bf B 461}
(1996) 539.
%
\bibitem{Jaffe} For a review see: R.L.\ Jaffe,
Talk given at Ettore Majorana International School of Nucleon Structure:
1st Course: The Spin Structure of the Nucleon, Erice, Italy, Aug.\ 3--10,
1995, MIT preprint MIT-CTP-2506, hep-ph/9602236.
%
\bibitem{AEL95}
M. Anselmino, A. Efremov, and E. Leader, Phys.\ Rep.\ {\bf 261}
(1995) 1.
%
\bibitem{Feynman} R.P.\ Feynman, in: Photon--Hadron Interactions,
Benjamin, 1972.
%
\bibitem{KR84} S.\ Kahana and G.\ Ripka,
Nucl.\ Phys.\ {\bf A 429} (1984) 462.
%
\bibitem{Witten}
E. Witten, Nucl.\ Phys.\ {\bf B 223} (1983) 433.
%
\bibitem{ANW}
G. Adkins, C. Nappi, and E. Witten, Nucl.\ Phys.\ {\bf B 228} (1983) 552.
%
\bibitem{DE}
D. Diakonov and M. Eides, Sov.\ Phys.\ JETP Lett.\ {\bf 38} (1983) 433.
%
\bibitem{DSW}
A. Dhar, R. Shankar, and S. Wadia, Phys.\ Rev.\ {\bf D 31} (1984) 3256.
%
\bibitem{DP86}
D. Diakonov and V. Petrov, Nucl.\ Phys.\ {\bf B 272} (1986) 457;
LNPI preprint LNPI-1153 (1986), published in: Hadron matter under extreme
conditions, Kiew (1986), p.192.
%
\bibitem{VJ90}
L. Vepstas and A.D.\ Jackson, Phys.\ Rep.\ {\bf 187} (1990) 109.
%
\bibitem{WGR}
H. Weigel, L. Gamberg, and H. Reinhardt, T\"ubingen University
preprint UNITU-THEP-6/1996, hep-ph/9604295; 
T\"ubingen University preprint 
UNITU-THEP-16/1996, hep-ph/9609226.
%
\bibitem{WW93} Ch.V.\ Christov {\em et al.}, Phys.\ Lett.\
{\bf B 325} (1994) 467; \\
M. Wakamatsu and T. Watabe, Phys.\ Lett.\ {\bf B 312} (1994) 184.
%
\end{thebibliography}
\end{document}